\documentclass[10pt,twocolumn]{IEEEtran}
\usepackage[latin1]{inputenc}
\usepackage{amsmath}
\usepackage{amsbsy}
\usepackage{amssymb}
\usepackage{latexsym}
\usepackage{times}
\usepackage{epsfig}

\title{\LARGE Adaptive Reduced-Rank LCMV Beamforming Algorithms Based on
Joint Iterative Optimization of Filters: Design and Analysis}

\author{R. C. de Lamare, L. Wang and R. Fa \\
\thanks{This work is partially funded by the
Ministry of Defense (MoD), UK, Project MoD, Contract No. RT/COM/S/021.
R. C. de Lamare, L. Wang and R. Fa are with the
Communications Research Group, Department of Electronics,
University of York, York Y010 5DD, United Kingdom . E-mails:
rcdl500@ohm.york.ac.uk} }

\begin{document}

\maketitle

\begin{abstract}
This paper presents reduced-rank linearly constrained minimum
variance (LCMV) beamforming algorithms based on joint iterative
optimization of filters. The proposed reduced-rank scheme is based
on a constrained joint iterative optimization of filters according
to the minimum variance criterion. The proposed optimization
procedure adjusts the parameters of a projection matrix and an
adaptive reduced-rank filter that operates at the output of the
bank of filters. We describe LCMV expressions for the design of
the projection matrix and the reduced-rank filter. We then
describe stochastic gradient and develop recursive least-squares
adaptive algorithms for their efficient implementation along with
automatic rank selection techniques. An analysis of the stability
and the convergence properties of the proposed algorithms is
presented and semi-analytical expressions are derived for
predicting their mean squared error (MSE) performance. Simulations
for a beamforming application show that the proposed scheme and
algorithms outperform in convergence and tracking the existing
full-rank and reduced-rank algorithms while requiring
comparable complexity.\\

\end{abstract}

\section{Introduction}

In recent years, adaptive beamforming techniques have attracted
considerable interest and found applications in radar, wireless
communications and sonar \cite{vantrees,li}. The adaptive
beamforming techniques are used in systems equipped with antenna
arrays and usually have a trade-off between performance and
computational complexity which depends on the designer's choice of
the adaptation algorithm \cite{haykin,veen,feldman}. The optimal
linearly constrained~ minimum~ variance (LCMV) beamformer is
designed in such a way that it attempts to minimize the array output
power while maintaining a constant response in the direction of a
signal of interest (SoI) \cite{vantrees,li,haykin}. However, this
technique requires the computation of the inverse of the input data
covariance matrix and the knowledge of the cross-correlation vector,
rendering the method very complex for practical applications when
the system is large. Adaptive versions of the LCMV beamformer were
subsequently reported with stochastic gradient (SG)
\cite{frost72,honig,delamaretsp} and recursive least squares (RLS)
\cite{romano96,honig} algorithms.

These algorithms require estimates of the input data covariance
matrix, which is a task that may become challenging in large
systems and in highly dynamic situations such as those found in
wireless communications and radar applications. This is because
the convergence speed and tracking properties of adaptive filters
depend on the number of sensor elements $M$ \cite{haykin} and on
the eigenvalue spread of the input data covariance matrix. Given
this dependency on the number of sensor elements $M$, it is thus
intuitive to reduce $M$ while simultaneously extracting the key
features of the original signal via an appropriate transformation.

A cost-effective technique in short-data record scenarios and, in
particular, with systems containing a large number of parameters
is reduced-rank signal processing. The advantages are their
superior convergence properties and enhanced tracking performance
when compared with full-rank schemes operating with a large number
of parameters, and their ability to exploit the low-rank nature of
the signals encountered in beamforming applications. Several
reduced-rank methods have been proposed to generate the signal
subspace \cite{haykin}-\cite{goldstein}. They range from
computationally expensive eigen-decomposition techniques
\cite{scharf}-\cite{bar-ness} to alternative approaches such as
the auxiliary-vector filter (AVF) \cite{pados99},\cite{karim02},
\cite{qian}, the multistage Wiener filter (MSWF) \cite{karim02},
\cite{reed98}, \cite{goldstein}, \cite{santos} which are based on
the Krylov subspace, and joint optimization approaches
\cite{hua,delamarespl07}. Despite the improved convergence and
tracking performance achieved with Krylov methods
\cite{pados99}-\cite{reed98}, \cite{goldstein}-\cite{qian} they
are relatively complex to implement and can suffer from numerical
problems. The joint optimization techniques reported in
\cite{hua,delamarespl07} outperform the eigen-decomposition- and
Krylov-based methods and are amenable to efficient adaptive
implementations. However, the design and analysis of adaptive LCMV
reduced-rank algorithms based on joint optimization approaches
have not been considered so far.

This work proposes LCMV reduced-rank algorithms based on
constrained joint iterative optimization of filters for
antenna-array beamforming. The proposed scheme, whose initial
results were reported in \cite{delamareel,delamareelb}, jointly
optimizes a projection matrix and a reduced-rank filter that
operates at the output of the projection matrix. The essence of
the proposed approach is to change the role of adaptive LCMV
filters. The bank of adaptive filters is responsible for
performing dimensionality reduction, whereas the reduced-rank
filter effectively forms the beam in the direction of the SoI. We
describe LCMV expressions for the design of the projection matrix
and the reduced-rank filter and present SG and RLS algorithms for
efficiently implementing the method. We also introduce an
automatic rank estimation algorithm for determining the most
adequate rank for the proposed algorithms. An analysis of the
stability and the convergence properties of the proposed
algorithms is presented and semi-analytical expressions are
derived for predicting their performance.

This paper is organized as follows. The system model is described
in Section II. The full-rank and the reduced-rank LCMV filtering
problems are formulated in Section III. Section IV is dedicated to
the proposed method, whereas Section V is devoted to the
derivation of the adaptive SG and RLS algorithms and the rank
adaptation technique. Section VI focuses on the analysis of the
proposed algorithms. Section VII presents and discusses the
simulation results and Section VIII gives the concluding remarks.

\section{System Model}

Let us consider a smart antenna system equipped with a uniform
linear array (ULA) of $M$ elements, as shown in Fig. 1. Assuming
that the sources are in the far field of the array, the signals of
$K$ narrowband sources impinge on the array $\left( K < M \right)$
with unknown directions of arrival (DOA) ${\theta}_l$ for $l=1,2,
\ldots, K$.

The input data from the antenna array can be organized in an $M
\times 1$ vector expressed by
\begin{equation}
{\boldsymbol r}(i) = {\boldsymbol A}(\theta) {\boldsymbol s}(i) +
{\boldsymbol n}(i)
\end{equation}
where $${\boldsymbol A}(\theta) = \left[ {\boldsymbol
a}(\theta_1), \ldots, {\boldsymbol a}(\theta_K)\right] $$ is the
$M \times K$ matrix of signal steering vectors.  The $M \times 1$
signal steering vector is defined as
\begin{equation}
{\boldsymbol a}(\theta_l) = \left[ 1, e^{-2\pi j
\frac{d_s}{\lambda_c}\cos \theta_l}, \ldots, e^{-2\pi j
(M-1)\frac{d_s}{\lambda_c}\cos \theta_l}\right] ^T
\end{equation}
for a signal impinging at angle $\theta_l$, $l=1,2, \ldots, K$,
where $d_s = \lambda_c / 2$ is the inter-element spacing,
$\lambda_c$ is the wavelength and $(.)^T$ denotes the transpose
operation. The vector ${\boldsymbol n}(i)$ denotes the complex
vector of sensor noise, which is assumed to be zero-mean and
Gaussian with covariance matrix $\sigma^2 {\boldsymbol I}$.

\section{Problem Statement}

In this section, we formulate the problems of full-rank and
reduced-rank LCMV filters. In order to perform beamforming with a
full-rank LCMV filter, we linearly combine the data vector
${\boldsymbol r}(i)=[r_{1}^{(i)}~r_{2}^{(i)}~ \ldots
~r_{M}^{(i)}]^{T}$ with the full-rank filter ${\boldsymbol w}=
[w_1^{}~ w_2^{} ~ \ldots ~ w_M^{}]^T$ to yield
\begin{equation}
x(i) = {\boldsymbol w}^{H} {\boldsymbol r}(i)
\end{equation}

The optimal LCMV filter is the $M \times 1$ vector ${\boldsymbol
w}$, which is designed to solve the following optimization problem
\begin{equation}
\begin{split}
{\rm minimize} ~  E[|{\boldsymbol w}^H {\bf r}(i) |^2] & = {\boldsymbol w}^{H} {\boldsymbol R} {\boldsymbol w}\\
{\rm subject~ to} ~ {\boldsymbol w}^H{\boldsymbol a}(\theta_k)~ &
=1 \label{flcmv}
\end{split}
\end{equation}
The solution to the problem in (\ref{flcmv}) is given by
\cite{haykin,frost72}
\begin{equation}
{\boldsymbol w}_{\rm opt} =  \frac{ {\boldsymbol
R}^{-1}{\boldsymbol a}(\theta_k)}{ {\boldsymbol a}^H(\theta_k)
{\boldsymbol R}^{-1} {\boldsymbol a}( \theta_k) \big)}
\end{equation}
where ${\boldsymbol a}(\theta_k)$ is the steering vector of the
SoI, ${\boldsymbol r}(i)$ is the received data, the covariance
matrix of ${\boldsymbol r}(i)$ is described by ${\boldsymbol
R}=E[{\boldsymbol r}(i){\boldsymbol r}^H(i)]$, $(\cdot)^{H}$
denotes Hermitian transpose and $E[\cdot]$ stands for expected
value. The filter ${\boldsymbol w}(i)$ can be estimated via SG or
RLS algorithms \cite{haykin}. However, the laws that govern their
convergence and tracking behaviors imply that they depend on $M$
and on the eigenvalue spread of ${\boldsymbol{R}}$.

A reduced-rank algorithm must extract the most important features
of the processed data by performing dimensionality reduction. This
mapping is carried out by a $M \times D$ projection matrix
${\boldsymbol S}_{D}$ on the received data as given by
\begin{equation}
\bar{\boldsymbol r}(i) = {\boldsymbol S}_D^H {\boldsymbol r}(i)
\end{equation}
where, in what follows, all $D$-dimensional quantities are denoted
with a "bar". The resulting projected received vector
$\bar{\boldsymbol r}(i)$ is the input to a filter represented by
the $D \time 1$ vector $\bar{\boldsymbol w}=[ \bar{w}_1^{}
~\bar{w}_2^{}~\ldots\bar{w}_D^{}]^T$. The filter output is
\begin{equation}
\bar{x}(i) = \bar{\boldsymbol w}^{H}\bar{\boldsymbol r}(i)
\end{equation}
In order to design the reduced-rank filter $\bar{\boldsymbol w}$
we consider the following optimization problem
\begin{equation}
\begin{split}
{\textrm{minimize}}  ~ E\big[ |\bar{\boldsymbol
w}^{H}\bar{\boldsymbol r}(i)|^2 \big] & = \bar{\boldsymbol
w}^{H} \bar{\boldsymbol R} \bar{\boldsymbol w} \\
{\textrm {subject to}}  ~ \bar{\boldsymbol w}^H\bar{\boldsymbol
a}(\theta_k) & = 1
\end{split}
\end{equation}
The solution to the above problem is
\begin{equation}
\bar{\boldsymbol w}_{\rm opt} =  \frac{ \bar{\boldsymbol
R}^{-1}\bar{\boldsymbol a}(\theta_k)}{ \bar{\boldsymbol
a}^H(\theta_k) \bar{\boldsymbol R}^{-1} \bar{\boldsymbol a}(
\theta_k)}
\end{equation}
where the reduced-rank covariance matrix is $\bar{\boldsymbol R} =
E[ \bar{\boldsymbol r}(i)\bar{\boldsymbol r}^{H}(i)]={\boldsymbol
S}_D^H{\boldsymbol R}{\boldsymbol S}_D$ and the reduced-rank
steering vector is $\bar{\boldsymbol a}(\theta_k)={\boldsymbol
S}_D^H {\boldsymbol a}(\theta_k)$. The associated minimum variance
(MV) for a LCMV filter with rank $D$ is
\begin{equation}
\begin{split}
{\rm MV} & = 
\frac{1}{ {\boldsymbol a}(\theta_k)^H{\boldsymbol S}_D
({\boldsymbol S}_D^H{\boldsymbol R}{\boldsymbol S}_D)^{-1}
{\boldsymbol S}_D^H {\boldsymbol a}(\theta_k)}
\end{split}
\end{equation}
The above development shows that the main problem is how to
cost-effectively design ${\boldsymbol S}_D$ to perform
dimensionality reduction on ${\boldsymbol r}(i)$, resulting in
improved convergence and tracking performance over the full-rank
filter. In the Appendix, we provide a necessary and sufficient
condition for ${\bf S}_D$ to preserve the MV of optimal full-rank
filter and discuss the existence of multiple solutions. In the
following, we detail our proposed reduced-rank method.

\section{Proposed Reduced-Rank Method}

In this section, we introduce the principles of the proposed
reduced-rank scheme. The proposed scheme, depicted in Fig. 2,
employs a matrix ${\boldsymbol S}_{D}(i)$ with dimensions $M
\times D$ to perform dimensionality reduction on a data vector
${\boldsymbol r}(i)$ with dimensions $M \times 1$. The
reduced-rank filter $\bar{\boldsymbol w}(i)$ with dimensions $D
\times 1$ processes the reduced-rank data vector $\bar{\boldsymbol
r}(i)$ in order to yield a scalar estimate $\bar{x}(i)$. The
projection matrix ${\bf S}_{D}(i)$ and the reduced-rank filter
$\bar{\boldsymbol w}(i)$ are jointly optimized in the proposed
scheme according to the MV criterion subject to a constraint that
ensures that the reduced-rank array response is equal to unity in
the direction of the SoI.

In order to describe the proposed method, let us first consider
the structure of the $M \times D$ projection matrix
\begin{equation}
{\boldsymbol S}_{D}(i) = [~{\boldsymbol s}_1(i) ~| ~{\boldsymbol
s}_2(i)~| ~\ldots~|{\boldsymbol s}_D(i)~]
\end{equation}
where the columns ${\boldsymbol s}_d(i)$ for $d = 1,~\ldots,~D$
constitute a bank of $D$ full-rank filters with dimensions $M
\times 1$ as given by
\begin{equation}
{\boldsymbol s}_d(i)=[s_{1,d}(i) ~ s_{2,d}(i)~ \ldots~s_{M,d}(i)
]^T \nonumber
\end{equation}
The output $\bar{x}(i)$ of the proposed reduced-rank scheme can be
expressed as a function of the input vector ${\bf r}(i)$, the
projection matrix ${\bf S}_D(i)$ and the reduced-rank filter
$\bar{\bf w}(i)$:
\begin{equation}
\begin{split}
\bar{x}(i) & =  \bar{\boldsymbol w}^H(i) {\boldsymbol S}_D^H(i)
{\boldsymbol r}(i) = \bar{\boldsymbol w}^H(i) \bar{\boldsymbol
r}(i)
\end{split}
\end{equation}
It is interesting to note that for $D=1$, the proposed scheme
becomes a conventional full-rank LCMV filtering scheme with an
addition weight parameter $w_D$ that provides an amplitude gain.
For $D>1$, the signal processing tasks are changed and the
full-rank LCMV filters compute a subspace projection and the
reduced-rank filter provides a unity gain in the direction of the
SoI. This rationale is fundamental to the exploitation of the
low-rank nature of signals in typical beamforming scenarios.

The LCMV expressions for the filters ${\bf S}_D(i)$ and $\bar{\bf
w}(i)$ can be computed via the proposed optimization problem
\begin{equation}
\begin{split}
{\textrm{minimize}}  ~ E\big[ |\bar{\boldsymbol
w}^{H}(i){\boldsymbol S}_D^H(i){\boldsymbol r}(i)|^2 \big] & =
\bar{\boldsymbol
w}^{H}(i) {\boldsymbol S}_D^H(i) {\boldsymbol R} {\boldsymbol S}_D(i) \bar{\boldsymbol w}(i) \\
{\textrm {subject to}}  ~ \bar{\boldsymbol w}^H(i){\boldsymbol
S}_D^H(i) {\boldsymbol a}(\theta_k) & = 1 \label{propt}
\end{split}
\end{equation}
In order to solve the above problem, we resort to the method of
Lagrange multipliers \cite{haykin} and transform the constrained
optimization into an unconstrained one expressed by the Lagrangian
\begin{equation}
\begin{split}
{\mathcal L}({\boldsymbol S}_D(i), \bar{\boldsymbol w}(i)) & =
E\big[ |\bar{\boldsymbol w}^{H}(i){\boldsymbol
S}_D^H(i){\boldsymbol r}(i)|^2 \big] + 2\Re [\lambda
(\bar{\boldsymbol w}^H(i){\bf S}_D^H(i){\boldsymbol
a}(\theta_k)-1) ], \label{uopt}
\end{split}
\end{equation}
where $\lambda$ is a scalar Lagrange multiplier, $*$ denotes
complex conjugate and the operator $\Re[ \cdot]$ selects the real
part of the argument. By fixing $\bar{\boldsymbol w}(i)$,
minimizing (\ref{uopt}) with respect to ${\boldsymbol S}_D(i)$ and
solving for $\lambda$, we get
\begin{equation}
\begin{split}
{\boldsymbol S}_D(i) & = \frac{ {\boldsymbol R}^{-1} {\boldsymbol
a}(\theta_k) \bar{\boldsymbol w}^H(i) \bar{\boldsymbol
R}_{\bar{w}}^{-1}}{\bar{\boldsymbol w}^H(i)\bar{\boldsymbol
R}_{\bar{w}}^{-1}\bar{\boldsymbol w}(i) {\boldsymbol
a}^H(\theta_k) {\boldsymbol R}^{-1} {\boldsymbol a}(\theta_k)},
\label{filts}
\end{split}
\end{equation}
where  ${\boldsymbol R} = E[{\boldsymbol r}(i){\boldsymbol
r}^{H}(i)]$ and $\bar{\boldsymbol R}_{\bar{w}} =
E[\bar{\boldsymbol w}(i)\bar{\boldsymbol w}^{H}(i)]$. By fixing
${\boldsymbol S}_D(i)$, minimizing (\ref{uopt}) with respect to
$\bar{\boldsymbol w}(i)$ and solving for $\lambda$, we arrive at
the expression
\begin{equation}
\bar{\boldsymbol w}(i) =  \frac{ \bar{\boldsymbol R}^{-1}(i)
\bar{\boldsymbol a}(\theta_k)}{\bar{\boldsymbol
a}^H(\theta_k)\bar{\boldsymbol R}^{-1}(i)\bar{\boldsymbol
a}(\theta_k)}, \label{filtw}
\end{equation}
where $\bar{\boldsymbol R}(i) = E[{\boldsymbol
S}_D^H(i){\boldsymbol r}(i){\boldsymbol r}^H(i)  {\boldsymbol
S}_D(i)] =E[\bar{\boldsymbol r}(i) \bar{\boldsymbol r}^{H}(i)]$,
$\bar{\boldsymbol a}(\theta_k) = {\boldsymbol
S}_D^H(i){\boldsymbol a}(\theta_k)$. The associated MV is
\begin{equation}
{\rm MV} = \frac{1}{ \bar{\boldsymbol a}^{H}(\theta_k)
\bar{\boldsymbol R}^{-1}(i) \bar{\boldsymbol a}(\theta_k) }.
\end{equation}
Note that the filter expressions in (\ref{filts}) and
(\ref{filtw}) are not closed-form solutions for $\bar{\boldsymbol
w}(i)$ and ${\boldsymbol S}_D(i)$ since (\ref{filts}) is a
function of $\bar{\boldsymbol w}(i)$ and (\ref{filtw}) depends on
${\boldsymbol S}_D(i)$. Thus, it is necessary to iterate
(\ref{filts}) and (\ref{filtw}) with initial values to obtain a
solution. An analysis of the optimization problem in (\ref{propt})
is given in Appendix II. Unlike existing approaches based on the
MSWF \cite{goldstein} and the AVF \cite{qian} methods, the
proposed scheme provides an iterative exchange of information
between the reduced-rank filter and the projection matrix and
leads to a much simpler adaptive implementation. The projection
matrix reduces the dimension of the input data, whereas the
reduced-rank filter yields a unity response in the direction of
the SoI. The key strategy lies in the joint optimization of the
filters. The rank $D$ must be set by the designer to ensure
appropriate performance or can be estimated via another algorithm.
In the next section, we seek iterative solutions via adaptive
algorithms for the design of ${\boldsymbol S}_D(i)$ and
$\bar{\boldsymbol w}(i)$, and automatic rank adaptation
algorithms.

\section{Adaptive Algorithms}
\label{sec:typestyle}

In this section we present adaptive SG and RLS versions of the
proposed scheme for efficient implementation. We also consider the
important issue of automatically determining the rank of the
scheme via the proposal of an adaptation technique. We then
provide the computational complexity in arithmetic operations of
the proposed reduced-rank algorithms.

\subsection{Stochastic Gradient Algorithm}

In this part, we present a low-complexity SG adaptive reduced-rank
algorithm for efficient implementation of the proposed method.
These algorithms were reported in \cite{delamareel,delamareelb}
and are reproduced here for convenience. By computing the
instantaneous gradient terms of (\ref{uopt}) with respect to
${\boldsymbol S}_D^*(i)$ and $\bar{\boldsymbol w}^*(i)$, we get
\begin{equation}
\begin{split}
\nabla {{\mathcal L}_{MV}}_{{\boldsymbol S}_D^*(i)} & =
\bar{x}^*(i) {\boldsymbol r}(i)\bar{\boldsymbol w}^H(i) + 2
\lambda^* {\bf a}(\theta_k) \bar{\boldsymbol w}^H(i) \\ \nabla
{{\mathcal L}_{MV}}_{\bar{\boldsymbol w}^*(i)} & = \bar{x}^*(i)
{\boldsymbol S}_D^H(i){\bf r}(i) + 2 \lambda^* {\boldsymbol
S}_D^H(i){\bf a}(\theta_k)
\end{split}
\end{equation}
By introducing the positive step sizes $\mu_s$ and $\mu_w$, using
the gradient rules ${\boldsymbol S}_D(i+1) = {\boldsymbol S}_D(i)
- \mu_s \nabla {{\mathcal L}_{MV}}_{{\boldsymbol S}_D^*(i)}$ and
$\bar{\boldsymbol w}(i+1) = \bar{\boldsymbol w}(i) - \mu_w \nabla
{{\mathcal L}_{MV}}_{\bar{\boldsymbol w}^*(i)}$, enforcing the
constraint and solving the resulting equations, we obtain
\begin{equation}
{\boldsymbol S}_D(i+1) = {\boldsymbol S}_D(i) - \mu_s \bar{x}^*(i)
\big[ {\boldsymbol r}(i)\bar{\boldsymbol w}^H(i) -
\big({\boldsymbol a}^H(\theta_k){\boldsymbol
a}(\theta_k)\big)^{-1}{\boldsymbol a}(\theta_k) \bar{\boldsymbol
w}^H(i) {\boldsymbol a}^H (\theta_k) {\boldsymbol r}(i)\big],
\label{recsd}
\end{equation}
\begin{equation}
\bar{\boldsymbol w}(i+1) = \bar{\boldsymbol w}(i) - \mu_w
\bar{x}^*(i) \big[{\boldsymbol I} - \big(\bar{\boldsymbol
a}^H(\theta_k)\bar{\boldsymbol a}(\theta_k)\big)^{-1}
\bar{\boldsymbol a}(\theta_k) \bar{\boldsymbol a}^H(\theta_k)
\big] \bar{\boldsymbol r}(i)\label{recw},
\end{equation}
where $\bar{x}(i)= \bar{\boldsymbol w}^H(i) {\boldsymbol S}_D^H(i)
{\boldsymbol r}(i)$. The proposed scheme trades-off a full-rank
filter against one projection matrix ${\boldsymbol S}_D(i)$ and
one reduced-rank adaptive filter $\bar{\boldsymbol w}(i)$
operating simultaneously and exchanging information.

\subsection{Recursive Least Squares Algorithms}

Here we derive an RLS adaptive reduced-rank algorithm for
efficient implementation of the proposed method. To this end, let
us first consider the Lagrangian
\begin{equation}
\begin{split}
\label{costls} {\mathcal{L}}_{\rm LS}({\boldsymbol S}_D(i),
\bar{\boldsymbol w}(i)) &
=\sum_{l=1}^{i}\alpha^{i-l}\big|\bar{\boldsymbol
w}^{H}(i){\boldsymbol S}_{D}^{H}(i)\boldsymbol r(l)\big|^{2} + 2
\Re [ \lambda\big(\bar{\boldsymbol w}^{H}(i) {\boldsymbol
S}_{D}^{H}(i)\boldsymbol a(\theta_k)-1\big) ]
\end{split}
\end{equation}
where $\alpha$ is the forgetting factor chosen as a positive
constant close to, but less than $1$.

Fixing $\bar{\boldsymbol w}(i)$, computing the gradient of
(\ref{costls}) with respect to $\boldsymbol S_{D}(i)$, equating
the gradient to a null vector and solving for $\lambda$, we obtain
\begin{equation}\label{filtsd}
\boldsymbol S_{D}(i)=\frac{\boldsymbol R^{-1}(i)\boldsymbol
a(\theta_k)\bar{\boldsymbol w}^{H}(i)\bar{\boldsymbol
R}_{\bar{w}}^{-1}(i)}{\bar{\boldsymbol w}^{H}(i)\bar{\boldsymbol
R}_{\bar{w}}^{-1}(i)\bar{\boldsymbol w}(i)\boldsymbol
a^{H}(\theta_k)\boldsymbol R^{-1}(i)\boldsymbol a(\theta_k)}
\end{equation}
where $\boldsymbol R(i)=\sum_{l=1}^{i}\alpha^{i-l}\boldsymbol
r(l)\boldsymbol r^{H}(l)$ is the input covariance matrix, and
$\bar{\boldsymbol R}_{\bar{w}}(i)=\bar{\boldsymbol
w}(i)\bar{\boldsymbol w}^{H}(i)$ is the reduced-rank weight matrix
at time instant $i$. The computation of (\ref{filtsd}) includes
the inversion of $\boldsymbol R(i)$ and $\bar{\boldsymbol
R}_{\bar{w}}(i)$, which may increase significantly the complexity
and create numerical problems. However, the expression in
(\ref{filtsd}) can be further simplified using the constraint
$\bar{\boldsymbol w}^{H}(i)\boldsymbol S_{D}^{H}(i)\boldsymbol
a(\theta_k)=1$. The details of the derivation of the proposed RLS
algorithms and the simplification are given in Appendix III. The
simplified expression for ${\boldsymbol S}_D(i)$ is given by
\begin{equation}\label{18}
\boldsymbol S_{D}(i)=\frac{\boldsymbol P(i)\boldsymbol
a(\theta_k)\bar{\boldsymbol a}^{H}(\theta_k)}{\boldsymbol
a^{H}(\theta_k)\boldsymbol P(i)\boldsymbol a(\theta_k)}
\end{equation}
where we defined the inverse covariance matrix ${\boldsymbol P}(i)
= \boldsymbol R^{-1}(i)$ for convenience of presentation.
Employing the matrix inversion lemma \cite{haykin}, we obtain
\begin{equation}\label{19}
\boldsymbol k(i)=\frac{\alpha^{-1}\boldsymbol P(i-1)\boldsymbol
r(i)}{1+\alpha^{-1}\boldsymbol r^{H}(i)\boldsymbol
P(i-1)\boldsymbol r(i)}
\end{equation}
\begin{equation}\label{20}
\boldsymbol P(i)=\alpha^{-1}\boldsymbol
P(i-1)-\alpha^{-1}\boldsymbol k(i)\boldsymbol r^{H}(i)\boldsymbol
P(i-1)
\end{equation}
where $\boldsymbol k(i)$ is the $M \times 1$ Kalman gain vector.
We set $\boldsymbol P(0)=\delta\boldsymbol I_M$ to start the
recursion of (\ref{20}), where $\delta$ is a positive constant and
$\boldsymbol I_M$ is an $M \times M$ identity matrix.

Assuming $\boldsymbol S_{D}(i)$ is known and taking the gradient
of (\ref{costls}) with respect to $\bar{\boldsymbol w}(i)$,
equating the terms to a null vector and solving for $\lambda$, we
obtain the $D \times 1$ reduced-rank filter
\begin{equation}\label{13}
\bar{\boldsymbol w}(i)=\frac{\bar{\boldsymbol
P}(i)\bar{\boldsymbol a}(\theta_k)}{\bar{\boldsymbol
a}^{H}(\theta_k)\bar{\boldsymbol P}(i)\bar{\boldsymbol
a}(\theta_k)}
\end{equation}
where $\bar{\boldsymbol P}(i) = \bar{\boldsymbol R}^{-1}(i)$ and
$\bar{\boldsymbol R}(i)=\sum_{l=1}^{i}\alpha^{i-l}\bar{\boldsymbol
r}(l)\bar{\boldsymbol r}^{H}(l)$ is the reduced-rank input
covariance matrix. In order to estimate $\bar{\boldsymbol P}(i)$,
we use the matrix inversion lemma \cite{haykin} as follows
\begin{equation}\label{15}
\bar{\boldsymbol k}(i)=\frac{\alpha^{-1}\bar{\boldsymbol
P}(i-1)\bar{\boldsymbol r}(i)}{1+\alpha^{-1}\bar{\boldsymbol
r}^{H}(i)\bar{\boldsymbol P}(i-1)\bar{\boldsymbol r}(i)}
\end{equation}
\begin{equation}\label{16}
\bar{\boldsymbol P}(i)=\alpha^{-1}\bar{\boldsymbol
P}(i-1)-\alpha^{-1}\bar{\boldsymbol k}(i)\bar{\boldsymbol
r}^{H}(i)\bar{\boldsymbol P}(i-1)
\end{equation}
where $\bar{\boldsymbol k}(i)$ is the $D \times 1$ reduced-rank
gain vector and $\bar{\boldsymbol P}(i)=\bar{\boldsymbol
R}^{-1}(i)$ is referred to as the reduced-rank inverse covariance
matrix. Hence, the covariance matrix inversion $\bar{\boldsymbol
R}^{-1}(i)$ is replaced at each step by the recursive processes
(\ref{15}) and (\ref{16}) for reducing the complexity.  The
recursion of (\ref{16}) is initialized by choosing
$\bar{\boldsymbol P}(0)=\bar{\delta}\bar{\boldsymbol I}_D$, where
$\bar{\delta}$ is a positive constant and $\bar{\boldsymbol I}_D$
is a $D \times D$ identity matrix.

The proposed RLS algorithm trade-off a full-rank filter with $M$
coefficients against one projection matrix ${\boldsymbol S}_D(i)$,
given in (\ref{18})-(\ref{20}) and one $ D \times 1$ reduced-rank
adaptive filter $\bar{\boldsymbol w}(i)$, given in
(\ref{13})-(\ref{16}), operating simultaneously and exchanging
information.

\subsection{Complexity of Proposed Algorithms}

Here, we evaluate the computational complexity of the proposed and
analyzed LCMV algorithms. The complexity expressed in terms of
additions and multiplications is depicted in Table I. We can
verify that the proposed reduced-rank SG algorithm has a
complexity that grows linearly with $DM$, which is about $D$ times
higher than the full-rank SG algorithm and significantly lower
than the MSWF-SG \cite{goldstein}. If $D << M$ (as we will see
later) then the additional complexity can be acceptable provided
the gains in performance justify them. In the case of the proposed
reduced-rank RLS algorithm the complexity is quadratic with $M^2$
and $D^2$. This corresponds to a complexity slightly higher than
the one observed for the full-rank RLS algorithm, provided $D$ is
significantly smaller than $M$, and comparable to the cost of the
MSWF-RLS \cite{goldstein} and the AVF \cite{qian}.

In order to illustrate the main trends in what concerns the
complexity of the proposed and analyzed algorithms, we show in
Fig. 3 the complexity in terms of additions and multiplications
versus the number of input samples $M$. The curves indicate that
the proposed reduced-rank RLS algorithm has a complexity lower
than the MSWF-RLS algorithm \cite{goldstein} and the AVF
\cite{qian}, whereas it remains at the same level of the full-rank
RLS algorithm. The proposed reduced-rank SG algorithm has a
complexity that is situated between the full-rank RLS and the
full-rank SG algorithms.

\subsection{Automatic Rank Selection }

The performance of the algorithms described in the previous
subsections depends on the rank $D$. This motivates the
development of methods to automatically adjust $D$ on the basis of
the cost function. Unlike prior methods for rank selection which
utilize MSWF-based algorithms \cite{goldstein} or AVF-based
recursions \cite{qian}, we focus on an approach that jointly
determines $D$ based on the LS criterion computed by the filters
${\boldsymbol S}_D(i)$ and $\bar{\boldsymbol w}_D(i)$, where the
subscript $D$ denotes the rank used for the adaptation. In
particular, we present a method for automatically selecting the
ranks of the algorithms based on the exponentially weighted
\textit{a posteriori} least-squares type cost function described
by
\begin{equation}
{\mathcal C}({\boldsymbol S}_D(i-1),\bar{\boldsymbol w}_D(i-1)) =
\sum_{l=1}^{i} \alpha^{i-l} \big|\bar{\boldsymbol
w}_D^{H}(i-1){\boldsymbol S}_D(i-1){\boldsymbol r}(l)|^2 ,
\label{eq:costadap}
\end{equation}
where $\alpha$ is the forgetting factor and $\bar{\bf w}_D(i-1)$
is the reduced-rank filter with rank $D$. For each time interval
$i$, we can select the rank $D_{\rm opt}$ which minimizes
${\mathcal C}({\boldsymbol S}_D(i-1),\bar{\boldsymbol
w}_{D}(i-1))$ and the exponential weighting factor $\alpha$ is
required as the optimal rank varies as a function of the data
record. The key quantities to be updated are the projection matrix
${\boldsymbol S}_D(i)$, the reduced-rank filter $\bar{\boldsymbol
w}_D(i)$, the associated reduced-rank steering vector
$\bar{\boldsymbol a}(\theta_k)$ and the inverse of the
reduced-rank covariance matrix $\bar{\boldsymbol P}(i)$ (for the
proposed RLS algorithm). To this end, we define the following
extended projection matrix $ {\boldsymbol S}_{D}(i)$ and the
extended reduced-rank filter weight vector $\bar{\boldsymbol
w}_{D}(i)$ as follows:
\begin{equation}
{\boldsymbol S}_{D}(i) = \left[\begin{array}{cccccc} s_{1,1} &
s_{1,2} & \ldots & s_{1,D_{\rm min}} & \ldots & s_{1,D_{\rm max}}
\\ \vdots & \vdots & \vdots & \vdots & \ddots & \vdots \\
s_{M,1} & s_{M,2} & \ldots & s_{M,D_{\rm min}} & \ldots &
s_{M,D_{\rm max}} \end{array} \right] ~~{\rm and} ~~
\bar{\boldsymbol w}_{D}(i) = \left[\begin{array}{c} w_1 \\ w_2
\\ \vdots \\ w_{D_{\rm min}} \\ \vdots \\ w_{D_{\rm max}} \end{array}\right]
\end{equation}
The extended projection matrix $ {\boldsymbol S}_{D}(i)$ and the
extended reduced-rank filter weight vector $\bar{\boldsymbol
w}_{D}(i)$ are updated along with the associated quantities
$\bar{\boldsymbol a}(\theta_k)$ and $\bar{\boldsymbol P}(i)$ (only
for the RLS) for the maximum allowed rank $D_{\rm max}$ and then
the proposed rank adaptation algorithm determines the rank that is
best for each time instant $i$ using the cost function in
(\ref{eq:costadap}). The proposed rank adaptation algorithm is
then given by
\begin{equation}
D_{\rm opt} = \arg \min_{D_{\rm min} \leq d \leq D_{\rm max}}
{\mathcal C}({\boldsymbol S}_D(i-1),\bar{\boldsymbol w}_D(i-1))
\end{equation}
where $d$ is an integer, $D_{\rm min}$ and $D_{\rm max}$ are the
minimum and maximum ranks allowed for the reduced-rank filter,
respectively. Note that a smaller rank may provide faster
adaptation during the initial stages of the estimation procedure
and a greater rank usually yields a better steady-state
performance. Our studies reveal that the range for which the rank
$D$ of the proposed algorithms have a positive impact on the
performance of the algorithms is limited, being from $D_{\rm
min}=3$ to $D_{\rm max}=8$ for the reduced-rank filter recursions.
These values are rather insensitive to the system load (number of
users), to the number of array elements and work very well for all
scenarios and algorithms examined. The additional complexity of
the proposed rank adaptation algorithm is that it requires the
update of all involved quantities with the maximum allowed rank
$D_{\rm max}$ and the computation of the cost function in
(\ref{eq:costadap}). This procedure can significantly improve the
convergence performance and can be relaxed (the rank can be made
fixed) once the algorithm reaches steady state. Choosing an
inadequate rank for adaptation may lead to performance
degradation, which gradually increases as the adaptation rank
deviates from the optimal rank. A mechanism for automatically
adjusting $D_{\rm min}$ and $D_{\rm max}$ based on a figure of
merit and the processed data would be an important technique to be
investigated. For example, this mechanism could in principle
adjust $D_{\rm min}$ and $D_{\rm max}$ in order to address the
needs of the model and the performance requirements. This remains
a topic for future investigation.

One can also argue that the proposed rank adaptation may not be
universally applied to signal processing problems, even though it
has been proven highly effective to the problems we dealt with.
Another possibility for rank adaptation is the use of the
cross-validation (CV) method reported in \cite{qian}. This
approach selects the lengths of the filters that minimize a cost
function that is estimated on the basis of data that have not been
used in the process of building the filters themselves. This
approach based on the concept of "leave one out" can be used to
determine the rank without requiring any prior knowledge or the
setting of a range of values \cite{qian}. A drawback of this
method is that it may significantly increase the length of the
filters, resulting in higher complexity. Other possible approaches
for rank selection may rely on some prior knowledge about the
environment and the system for inferring the required rank for
operation. The development of cost-effective methods for rank
selection remains an interesting area for investigation.

\section{Analysis of Algorithms}

In this section, we present the stability and the MSE convergence
analyses of the proposed SG algorithms.  Specifically, we consider
the joint optimization approach and derive conditions of stability
for the proposed SG algorithms. We then assume that the algorithms
will converge and carry out the MSE convergence analysis in order
to semi-analytically determine the MSE upon convergence. The RLS
algorithms are expected to converge to the optimal LCMV filter and
this has been verified in our studies. A discussion on the
preservation of the MV performance, the existence of multiple
solutions and an analysis of the optimization of the proposed
scheme valid for both SG and RLS algorithms is included in the
Appendices I and II.

\subsection{Stability Analysis}

In order to establish conditions for the stability of the proposed
SG algorithms, we define the error matrices at time $i$ as
$${\boldsymbol e}_{{\boldsymbol S}_D}(i) = {\boldsymbol S}_D(i) - {\boldsymbol
S}_{D,{\rm opt}}$$ and $${\boldsymbol e}_{\bar{\boldsymbol w}}(i)
= \bar{\boldsymbol w}(i) - \bar{\boldsymbol w}_{\rm opt},$$ where
$\bar{\boldsymbol w}_{\rm opt}$ and ${\boldsymbol S}_{D,{\rm
opt}}$ are the optimal parameter estimators. Since we are dealing
with a joint optimization procedure, both filters have to be
considered jointly. By substituting the expressions of
${\boldsymbol e}_{{\boldsymbol S}_D}(i)$ and ${\boldsymbol
e}_{\bar{\boldsymbol w}}(i)$ in (\ref{recsd}) and (\ref{recw}),
respectively, and rearranging the terms we obtain {
\begin{equation}
\begin{split}
{\boldsymbol e}_{{\boldsymbol S}_D}(i+1) & = \big\{ {\boldsymbol
I} - \mu_s [{\boldsymbol I} - ({\boldsymbol
a}^H(\theta_k){\boldsymbol a}(\theta_k) )^{-1}{\boldsymbol
a}(\theta_k) {\boldsymbol a}^H(\theta_k)] {\boldsymbol r}(i)
{\boldsymbol r}^H(i) \big\} {\boldsymbol e}_{{\boldsymbol S}_D}(i)
\\ & \quad - \mu_s [{\boldsymbol I} - ({\boldsymbol
a}^H(\theta_k){\boldsymbol a}(\theta_k) )^{-1}{\boldsymbol
a}(\theta_k) {\boldsymbol a}^H (\theta_k)] {\boldsymbol r}(i)
\bar{\boldsymbol w}^H(i) {\boldsymbol r}^H(i){\boldsymbol S}_D(i)
{\boldsymbol e}_{\bar{\boldsymbol w}}(i) \\ & \quad + \mu_s
[{\boldsymbol I} - ({\boldsymbol a}^H(\theta_k){\boldsymbol
a}(\theta_k) )^{-1}{\boldsymbol a}(\theta_k) {\boldsymbol a}^H
(\theta_k)] {\boldsymbol r}(i) {\boldsymbol r}^H(i) [ {\boldsymbol
S}_D(i)({\bf I} - \bar{\boldsymbol w}_{\rm opt} \bar{\boldsymbol
w}^H(i)) - {\boldsymbol S}_{D,{\rm opt}} ] \label{esd}
\end{split}
\end{equation}
\begin{equation}
\begin{split}
{\boldsymbol e}_{{\boldsymbol w}}(i+1) & = \big\{ {\boldsymbol I}
- \mu_w [{\boldsymbol I} - (\bar{\boldsymbol a}^H(\theta_k)
\bar{\boldsymbol a}(\theta_k))^{-1}\bar{\boldsymbol a}(\theta_k)
\bar{\boldsymbol a}^H(\theta_k)] \bar{\boldsymbol r}(i)
\bar{\boldsymbol r}^H(i) \big\} {\boldsymbol e}_{{\boldsymbol
w}}(i)
\\ &  \quad - \mu_w [{\boldsymbol I} - (\bar{\boldsymbol a}^H(\theta_k)
\bar{\boldsymbol a}(\theta_k))^{-1}\bar{\boldsymbol a}(\theta_k)
\bar{\boldsymbol a}^H(\theta_k)] \bar{\boldsymbol r}(i)
{\boldsymbol r}^H(i)  {\boldsymbol e}_{{\boldsymbol S}_D}(i)
\\ &  \quad + \mu_w [{\boldsymbol I} - (\bar{\boldsymbol a}^H(\theta_k)
\bar{\boldsymbol a}(\theta_k))^{-1}\bar{\boldsymbol a}(\theta_k)
\bar{\boldsymbol a}^H(\theta_k)){\boldsymbol S}_{D}^H(i)]
\bar{\boldsymbol r}(i) \bar{\boldsymbol r}^H(i) ( {\boldsymbol
S}_{D}(i)( {\boldsymbol I}- \bar{\boldsymbol w}_{\rm opt}) -
{\boldsymbol S}_{D,{\rm opt}}) \label{ew}
\end{split}
\end{equation}}

Taking expectations and simplifying the terms, we obtain
\begin{equation}
\begin{split}
\left[\begin{array}{c}
  E[{\boldsymbol e}_{{\boldsymbol S}_D}(i+1)] \\
  E[{\boldsymbol e}_{\bar{\boldsymbol w}}(i+1)]
\end{array}\right] & = {\boldsymbol P}
\left[\begin{array}{c}
  E[{\boldsymbol e}_{{\boldsymbol S}_D}(i)] \\
  E[{\boldsymbol e}_{\bar{\boldsymbol w}}(i)]
\end{array}\right] + {\boldsymbol T} \\
 \end{split}
\end{equation}
where \begin{equation}{\boldsymbol P} =  \left[\begin{array}{c c}
  {\small \big\{ {\boldsymbol
I} - \mu_s [{\boldsymbol I} - ({\boldsymbol
a}^H(\theta_k){\boldsymbol a}(\theta_k) )^{-1}{\boldsymbol
a}(\theta_k) {\boldsymbol a}^H(\theta_k)] {\boldsymbol r}(i)
{\boldsymbol r}^H(i) \big\}} & {\small - \mu_s [{\boldsymbol I} -
{\boldsymbol a}(\theta_k) {\boldsymbol a}^H (\theta_k)]
{\boldsymbol r}(i) \bar{\boldsymbol w}^H(i)
{\boldsymbol r}^H(i){\boldsymbol S}_D(i)} \\
  {\small - \mu_w [{\boldsymbol I} - (\bar{\boldsymbol a}^H(\theta_k)
\bar{\boldsymbol a}(\theta_k))^{-1}\bar{\boldsymbol a}(\theta_k)
\bar{\boldsymbol a}^H(\theta_k)] \bar{\boldsymbol r}(i)
{\boldsymbol r}^H(i) }  & {\small \big\{ {\boldsymbol I} - \mu_w
[{\boldsymbol I} - (\bar{\boldsymbol a}^H(\theta_k)
\bar{\boldsymbol a}(\theta_k))^{-1}\bar{\boldsymbol a}(\theta_k)
\bar{\boldsymbol a}^H(\theta_k)] \bar{\boldsymbol r}(i)
\bar{\boldsymbol r}^H(i) \big\}}
\end{array}\right],\nonumber\end{equation}
\begin{equation}{\boldsymbol T} = \left[\begin{array}{c}
{\small  \mu_s [{\boldsymbol I} - ({\boldsymbol
a}^H(\theta_k){\boldsymbol a}(\theta_k) )^{-1}{\boldsymbol
a}(\theta_k) {\boldsymbol a}^H (\theta_k)] {\boldsymbol r}(i)
{\boldsymbol r}^H(i) [ {\boldsymbol S}_D(i)({\bf I} -
\bar{\boldsymbol w}_{\rm opt} \bar{\boldsymbol
w}^H(i)) - {\boldsymbol S}_{D,{\rm opt}} ]} \\
{\small  \mu_w [{\boldsymbol I} - (\bar{\boldsymbol a}^H(\theta_k)
\bar{\boldsymbol a}(\theta_k))^{-1}\bar{\boldsymbol a}(\theta_k)
\bar{\boldsymbol a}^H(\theta_k)){\boldsymbol S}_{D}^H(i)]
\bar{\boldsymbol r}(i) \bar{\boldsymbol r}^H(i) ( {\boldsymbol
S}_{D}(i)( {\boldsymbol I}- \bar{\boldsymbol w}_{\rm opt}) -
{\boldsymbol S}_{D,{\rm opt}}) }
\end{array}\right].\nonumber
\end{equation}
The previous equations imply that the stability of the algorithms
depends on the spectral radius of ${\boldsymbol P}$. For
convergence, the step sizes should be chosen such the eigenvalues
of ${\boldsymbol P}^H{\boldsymbol P}$ are less than one. Unlike
the stability analysis of most adaptive algorithms \cite{haykin},
in the proposed approach the terms are more involved and depend on
each other as evidenced by the equations in ${\boldsymbol P}$ and
${\boldsymbol T}$.

\subsection{MSE Convergence Analysis}

Let us consider in this part an analysis of the MSE in steady
state. This follows the general steps of the MSE convergence
analysis of \cite{haykin} even though novel elements will be
introduced in the proposed framework. These novel elements in the
analysis are the joint optimization of the two adaptive filters
$\bar{\boldsymbol w}(i)$ and ${\boldsymbol S}_D(i)$ of the
proposed scheme and a strategy to incorporate the effect of the
step size of the recursions in (\ref{recsd}) and (\ref{recw}).

Let us define the MSE at time $i+1$ using the relations
$${\boldsymbol e}_{\boldsymbol w}(i+1) = {\boldsymbol w}(i+1) - {\boldsymbol
w}_{opt}$$ and
$$\xi (i) = E[ {\boldsymbol w}^H(i){\boldsymbol r}(i) {\boldsymbol
r}(i) {\boldsymbol w}(i)],$$ where the filter ${\boldsymbol w}(i)
= {\boldsymbol S}_D(i) \bar{\boldsymbol w}(i)$ with $M$
coefficients is the $D$-rank approximation of a full-rank filter
obtained with an inverse mapping performed by ${\boldsymbol
S}_D(i)$.

The MSE of the proposed scheme can be expressed by:
\begin{equation}
\begin{split}
{\rm MSE}(i) &  =  E[|d(i) - {\boldsymbol w}^{H}(i){\boldsymbol r}(i)|^{2}] \\
& = \epsilon_{min} + \xi(i) - \xi_{min} - E[{\boldsymbol
e}^{H}_{\boldsymbol w}(i)] {\boldsymbol a}(\theta_k) -
{\boldsymbol a}^{H}({\theta_k})E[{\boldsymbol e}_{\boldsymbol w}(i)]\\
& = \epsilon_{min} + \xi_{ex}(i) - E[{\boldsymbol
e}^{H}_{\boldsymbol w}(i)] {\boldsymbol a}({\theta_k}) -
{\boldsymbol a}^{H}(\theta_{k})E[{\boldsymbol e}_{\boldsymbol
w}(i)]
\end{split}
\end{equation}
where $d(i)$ corresponds to the desired signal, $\xi(i) =
E[{\boldsymbol w}^{H}(i){\boldsymbol r}(i){\boldsymbol
r}^{H}(i){\boldsymbol w}(i)]$, $\epsilon_{min} =
E[|d(i)-{\boldsymbol w}^{H}_{opt}{\boldsymbol r}(i)|^{2}]$ is the
MSE with
\begin{equation}
{\boldsymbol w}_{opt} = \xi_{min}{\boldsymbol R}^{-1}{\boldsymbol
a}({\theta_k}), \label{optf}
\end{equation}
where $\xi_{min}=1/({\boldsymbol a}^{H}({\theta_k}){\boldsymbol
R}^{-1}{\boldsymbol a}({\theta_k}))$ is the minimum variance, and
$\xi_{ex}(i) = \xi(i) - \xi_{min}$ is the excess MSE due to the
adaptation process at the time instant $i$. Since
$\lim_{i\rightarrow \infty} E[{\boldsymbol e}_{\boldsymbol w}(i)]
= 0$ we have
\begin{equation}
\lim_{i\rightarrow \infty} {\rm MSE}(i) = \epsilon_{min} +
\lim_{i\rightarrow \infty} \xi_{ex}(i) \label{exmse}
\end{equation}
where the $\xi_{ex}(\infty)$ term in (\ref{exmse}) is the
steady-state excess MSE resulting from the adaptation process. The
main difference here from prior work lies in the fact that this
refers to the excess MSE produced by a $D$-rank approximation
filter ${\boldsymbol w}(i)$. In order to analyze the trajectory of
$\xi(i)$, let us rewrite it as
\begin{equation}
\begin{split}
\xi(i) & = E[{\boldsymbol w}^{H}(i){\boldsymbol r}(i){\boldsymbol
r}^{H}(i){\boldsymbol w}(i)] \\ & = E[ \bar{\boldsymbol w}^H(i)
{\boldsymbol S}_D^H(i) {\boldsymbol r}(i) {\boldsymbol r}^H(i)
{\boldsymbol S}_D(i) \bar{\boldsymbol w}(i)] \\ & =
tr~E[{\boldsymbol R}_{\boldsymbol w }(i){\boldsymbol R}]
\label{exmv}
\end{split}
\end{equation}
where ${\boldsymbol R}_{\boldsymbol w}(i)= E[ {\boldsymbol
w}(i){\boldsymbol w}^H(i)] = {\boldsymbol w}_{opt}{\boldsymbol
w}_{opt}^{H} + E[{\boldsymbol e}_{w}(i)]{\boldsymbol w}_{opt}^{H}
+ {\boldsymbol w}_{opt}E[{\boldsymbol e}_{\boldsymbol w}^{H}(i)] +
{\boldsymbol R}_{e_{\boldsymbol w}}(i)$ \cite{honig}.

To proceed with the analysis, we must define the quantities
${\boldsymbol R}=\boldsymbol{\Phi} \boldsymbol{\Lambda}
\boldsymbol{\Phi}^{H}$, where the columns of $\boldsymbol{\Phi}$
are the eigenvectors of the symmetric and positive semi-definite
matrix ${\boldsymbol R}$ and $\boldsymbol{\Lambda}$ is the
diagonal matrix of the corresponding eigenvalues, $ {\boldsymbol
R}_{{\boldsymbol e}_w}(i) = E[{\boldsymbol e}_{\boldsymbol w}(i)
{\boldsymbol e}_{\boldsymbol w}^{H}(i)]$, the rotated tap error
vector $\tilde{{\boldsymbol e}}_{\boldsymbol
w}(i)=\boldsymbol{\Phi}^{H}{\boldsymbol e}_{\boldsymbol w}(i)$,
the rotated signal vectors $\tilde{\boldsymbol
r}(i)=\boldsymbol{\Phi}^{H}{\boldsymbol r}(i)$,
$\tilde{\boldsymbol a}({\theta_k})
=\boldsymbol{\Phi}^{H}{\boldsymbol a}({\theta_k})$ and
${\boldsymbol R}_{\tilde{\boldsymbol e}_{\boldsymbol w}}(i) =
E[\tilde{\boldsymbol e}_{\boldsymbol w}(i)\tilde{\boldsymbol
e}_{\boldsymbol w}^{H}(i)]=\boldsymbol{\Phi}^{H}{\boldsymbol
R}_{{\boldsymbol e}_{\boldsymbol w}}(i)\boldsymbol{\Phi}$.
Rewriting (\ref{exmv}) in terms of the above transformed
quantities we have:
\begin{equation}
\begin{split}
\xi(i) & = tr ~E[\boldsymbol{\Lambda} \boldsymbol{\Phi}^{H}
{\boldsymbol R}_{\boldsymbol w} \boldsymbol{\Phi}] \\ & =
\xi_{min} + tr[ E[\tilde{\boldsymbol e}_{\boldsymbol w}(i)]
\tilde{\boldsymbol a}^{H}({\theta_k}) + \tilde{\boldsymbol
a}({\theta_k}) E[\tilde{\boldsymbol e}_{\boldsymbol w}^{H}(i)] \\
& \quad + \Lambda{\boldsymbol R}_{\tilde{\boldsymbol
e}_{\boldsymbol w}}(i)]
\end{split}
\end{equation}
Since $\lim_{i\rightarrow \infty} E[\tilde{\boldsymbol
e}_{\boldsymbol w}(i)]=0$, then $\lim_{i\rightarrow \infty} \xi(i)
= \xi_{min} + tr[ \Lambda{\boldsymbol R}_{\tilde{\boldsymbol
e}_{\boldsymbol w}}]$. Thus, it is evident that to assess the
evolution of $\xi(i)$ it is sufficient to study ${\boldsymbol
R}_{\tilde{\boldsymbol e}_{\boldsymbol w}}(i)$.

Using ${\boldsymbol e}_{{\boldsymbol S}_D}(i)$ and ${\boldsymbol
e}_{\bar{\boldsymbol w}}(i)$ and combining them to compute
${\boldsymbol e}_{\boldsymbol w}(i)$, we get
\begin{equation}
\begin{split}
{\boldsymbol e}_{\boldsymbol w}(i) & = {\boldsymbol w}(i) -
{\boldsymbol w}_{\rm opt} \\ & = {\boldsymbol S}_D(i)
\bar{\boldsymbol w}(i) - {\boldsymbol S}_{D,~{\rm opt}} \bar{\boldsymbol w}_{\rm opt} \\
& = {\boldsymbol e}_{{\boldsymbol S}_D}(i) {\boldsymbol
e}_{\bar{\boldsymbol w}}(i) + {\boldsymbol S}_{D, {\rm opt}}
{\boldsymbol e}_{\bar{\boldsymbol w}}(i) + {\boldsymbol
e}_{{\boldsymbol S}_D}(i) \bar{\boldsymbol w}_{\rm opt}
\end{split}
\end{equation}
Substituting the expressions for ${\boldsymbol e}_{{\boldsymbol
S}_D}(i+1)$ and ${\boldsymbol e}_{\bar{\boldsymbol w}}(i+1)$ in
(\ref{esd}) and (\ref{ew}), respectively, to compute ${\boldsymbol
e}_{\boldsymbol w}(i+1)$, we get
\begin{equation}
\begin{split}
{\boldsymbol e}_{\boldsymbol w}(i+1) & = {\boldsymbol
e}_{\boldsymbol w}(i) - \mu_w \bar{x}^*(i) {\boldsymbol S}_D(i)
\bar{\boldsymbol r}_p(i) - \mu_s \bar{x}^*(i) {\boldsymbol
S}_{{\boldsymbol r}_p}(i) \bar{\boldsymbol w}(i) \\ & \quad +
\mu_s \mu_w (\bar{x}^*(i))^2 {\boldsymbol S}_{{\boldsymbol
r}_p}(i) \bar{\boldsymbol r}_p (i) + {\boldsymbol S}_{D, {\rm
opt}} {\boldsymbol e}_{\bar{\boldsymbol w}}(i) + {\boldsymbol
e}_{{\boldsymbol S}_D}(i) \bar{\boldsymbol w}_{opt} \label{ewg}
\end{split}
\end{equation}
where
$$\bar{x}(i) = \bar{\boldsymbol w}^H(i) {\boldsymbol S}_{D}^H(i)
{\boldsymbol r}(i) = {\boldsymbol w}^H(i) {\boldsymbol r}(i)$$
$${\boldsymbol S}_{{\boldsymbol r}_p}(i) = \big({\boldsymbol I} - ({\boldsymbol
a}^H(\theta_k){\boldsymbol a}(\theta_k) )^{-1}{\boldsymbol
a}(\theta_k)
 {\boldsymbol a}^H(\theta_k)\big) {\boldsymbol
r}(i)\bar{\boldsymbol w}^H(i)$$
$$ \bar{\boldsymbol r}_p (i) = ({\boldsymbol I} - ({\boldsymbol
S}_D(i) {\boldsymbol a}^H(\theta_k) {\boldsymbol S}_D^H(i)
{\boldsymbol a}(\theta_k))^{-1} {\boldsymbol S}_D(i) {\boldsymbol
a}(\theta_k) {\boldsymbol a}^H(\theta_k) {\boldsymbol S}_D^H(i)
{\boldsymbol S}_D(i) {\boldsymbol r}(i)$$ We can further rewrite
the expressions above in order to obtain a more compact and
convenient representation as
\begin{equation}
\begin{split}
{\boldsymbol e}_{\boldsymbol w}(i+1) & =  
(I - {\boldsymbol A})
{\boldsymbol e}_{\boldsymbol w}(i) + {\boldsymbol B}{\boldsymbol
C} + \mu_s \mu_w (\bar{x}^*(i))^2 {\boldsymbol S}_{{\boldsymbol
r}_p}(i) \bar{\boldsymbol r}_p(i) + {\boldsymbol e}_{{\boldsymbol
S}_D}(i) \bar{\boldsymbol w}_{opt} \label{ewg2}
\end{split}
\end{equation}
where $${\boldsymbol A} =   \mu_w {\boldsymbol S}_D(i)
\bar{\boldsymbol r}_p(i) {\boldsymbol r}^H(i) + \mu_s {\boldsymbol
S}_{{\boldsymbol r}_p}(i) \bar{\boldsymbol w}(i) {\boldsymbol
r}^H(i)-{\boldsymbol S}_{D, {\rm opt}}$$ $${\boldsymbol B} = -
\mu_w {\boldsymbol S}_D(i) \bar{\boldsymbol r}_p(i) {\boldsymbol
r}^H(i) - \mu_s {\boldsymbol S}_{{\boldsymbol r}_p}(i)
\bar{\boldsymbol w}(i) {\boldsymbol r}^H(i)$$
$${\boldsymbol C} = {\boldsymbol e}_{{\boldsymbol S}_D}(i)
\bar{\boldsymbol w}_{\rm opt} + {\boldsymbol S}_{D, {\rm opt}}
{\boldsymbol e}_{\bar{\boldsymbol w}}(i) + {\boldsymbol
e}_{{\boldsymbol S}_D}(i) \bar{\boldsymbol w}_{opt}.$$

Now, we need to compute ${\boldsymbol R}_{{\boldsymbol
e}_{\boldsymbol w}}(i+1) = E[{\boldsymbol e}_{\boldsymbol w}(i+1)
{\boldsymbol e}_{\boldsymbol w}^H(i+1)]$ by using the result in
(\ref{ewg2}), which yields
\begin{equation}
\begin{split}
{\boldsymbol R}_{{\boldsymbol e}_{\boldsymbol w}}(i+1) & =
({\boldsymbol I} - {\boldsymbol A}){\boldsymbol R}_{{\boldsymbol
e}_{\boldsymbol w}}(i) ({\boldsymbol I} - {\boldsymbol A})^H +
({\boldsymbol I} - {\boldsymbol A}){\boldsymbol e}_{\boldsymbol
w}(i) {\boldsymbol C}^H {\boldsymbol B}^H \\ &  \quad + \mu_s
\mu_w (\bar{x}(i))^2 ({\boldsymbol I} - {\boldsymbol
A}){\boldsymbol e}_{\boldsymbol w}(i) (\bar{\boldsymbol r}_p^H(i)
{\boldsymbol S}_{{\boldsymbol r}_p}^H(i) )  \\ & \quad +
({\boldsymbol I}- {\boldsymbol A}) {\boldsymbol e}_{\boldsymbol
w}(i) \bar{\boldsymbol w}_{\rm opt}^H {\boldsymbol S}_{D,{\rm
opt}}^H + {\boldsymbol B}{\boldsymbol C} {\boldsymbol
e}_{\boldsymbol w}^H(i) (({\boldsymbol I} - {\boldsymbol A})^H \\
& \quad + {\boldsymbol B}{\boldsymbol C} {\boldsymbol C}^H
{\boldsymbol B}^H + \mu_s \mu_w (\bar{x}(i))^2 {\boldsymbol B}
{\boldsymbol C} \bar{\boldsymbol r}_p^H(i){\boldsymbol
S}_{{\boldsymbol r}_p}^H(i)
\\ & \quad + {\boldsymbol B}{\boldsymbol C} {\boldsymbol w}_{\rm opt}^H
{\boldsymbol e}_{{\boldsymbol S}_D}^H(i) + \mu_s \mu_w
(\bar{x}^*(i))^2 {\boldsymbol S}_{ {\boldsymbol r}_p}(i)
{\boldsymbol r}_p(i)
{\boldsymbol e}_{\boldsymbol w}^H(i) ({\boldsymbol I} - {\boldsymbol A})^H \\
& \quad + \mu_s \mu_w (\bar{x}^*(i))^2 {\boldsymbol
S}_{{\boldsymbol r}_p}(i) \bar{\boldsymbol r}_p(i) {\boldsymbol
C}^H {\boldsymbol A}^H
\\ & \quad + ( \mu_s \mu_w)^2 |\bar{x}(i)|^4 {\boldsymbol
S}_{{\boldsymbol r}_p}(i) \bar{\boldsymbol r}_p(i)
\bar{\boldsymbol r}_p^H(i) {\boldsymbol S}_{{\boldsymbol r}_p}(i)
\\ & \quad + \mu_s \mu_w (\bar{x}(i))^2 {\boldsymbol
e}_{{\boldsymbol S}_D}(i) \bar{\boldsymbol w}_{\rm opt}
\bar{\boldsymbol r}_p(i) {\boldsymbol S}_{{\boldsymbol r}_p}(i) \\
& \quad - {\boldsymbol e}_{{\boldsymbol S}_D}(i) \bar{\boldsymbol
w}_{opt} {\boldsymbol e}^H_{\boldsymbol w}(i) ({\boldsymbol I} -
{\boldsymbol A})^H  + {\boldsymbol e}_{{\boldsymbol S}_D}(i)
\bar{\boldsymbol w}_{opt} {\boldsymbol C}^H {\boldsymbol B}^H \\
 & \quad + {\boldsymbol
e}_{{\boldsymbol S}_D}(i) \bar{\boldsymbol w}_{\rm opt}
\bar{\boldsymbol w}_{\rm opt}^H {\boldsymbol e}_{{\boldsymbol
S}_D}^H(i) \label{mre}
\end{split}
\end{equation}
Since $E[{\boldsymbol e}_{{\boldsymbol w}}(i)]={\boldsymbol 0}$
and $E[{\boldsymbol e}_{{\boldsymbol S}_D}(i)] = {\boldsymbol 0}$,
we can simplify the previous expression and obtain
\begin{equation}
\begin{split}
{\boldsymbol R}_{{\boldsymbol e}_{\boldsymbol w}}(i+1) & =
({\boldsymbol I} - {\boldsymbol A}){\boldsymbol R}_{{\boldsymbol
e}_{\boldsymbol w}}(i) ({\boldsymbol I} - {\boldsymbol A})^H \\ &
\quad  + {\boldsymbol B}{\boldsymbol C} {\boldsymbol C}^H
{\boldsymbol B}^H + \mu_s \mu_w (\bar{x}(i))^2 {\boldsymbol B}
{\boldsymbol C} \bar{\boldsymbol r}_p^H(i){\boldsymbol
S}_{{\boldsymbol r}_p}^H(i)
\\ & \quad    + \mu_s \mu_w (\bar{x}^*(i))^2 {\boldsymbol S}_{{\boldsymbol
r}_p}(i) \bar{\boldsymbol r}_p(i) {\boldsymbol C}^H {\boldsymbol
A}^H
\\ & \quad + ( \mu_s \mu_w)^2 |\bar{x}(i)|^4 {\boldsymbol
S}_{{\boldsymbol r}_p}(i) \bar{\boldsymbol r}_p(i)
\bar{\boldsymbol r}_p^H(i) {\boldsymbol S}_{{\boldsymbol r}_p}(i)
\\ & \quad  + {\boldsymbol
e}_{{\boldsymbol S}_D}(i) \bar{\boldsymbol w}_{\rm opt}
\bar{\boldsymbol w}_{\rm opt}^H {\boldsymbol e}_{{\boldsymbol
S}_D}^H(i) \label{mresimp}
\end{split}
\end{equation}
Solving for ${\boldsymbol R}_{{\boldsymbol e}_{\boldsymbol w}}$,
the MSE can be computed by
\begin{equation}
\begin{split}
{\rm MSE}(i+1) &  =  \epsilon_{min} + tr[  \Lambda{\boldsymbol
R}_{\tilde{\boldsymbol e}_{{\boldsymbol w}}}(i)] \\ & =
\epsilon_{min} + tr[  \Lambda {\boldsymbol \Phi} {\boldsymbol
R}_{{\boldsymbol e}_{{\boldsymbol w}}}(i){\boldsymbol \Phi}^H]
\label{fmse}
\end{split}
\end{equation}
It should be remarked that the expression for ${\boldsymbol
R}_{{\boldsymbol e}_{\boldsymbol w}}(i)$ is quite involved and
requires a semi-analytical approach with the aid of computer
simulations for its computation. This is because the terms
resulting from the joint adaptation create numerous extra terms in
the expression of ${\boldsymbol R}_{{\boldsymbol e}_{\boldsymbol
w}}(i)$, which are very difficult to isolate. We found that using
computer simulations to pre-compute the terms of ${\boldsymbol
R}_{{\boldsymbol e}_{\boldsymbol w}}(i)$ as a function of the step
sizes was more practical and resulted in good match between the
semi-analytical and simulated curves. In the following section, we
will demonstrate that it is able to predict the performance of the
proposed SG algorithm.

\section{Simulations}

In this section we evaluate the performance of the proposed and
the analyzed beamforming algorithms via computer simulations. We
also verify the validity of the MSE convergence analysis of the
previous section. A smart antenna system with a ULA containing $M$
sensor elements is considered for assessing the beamforming
algorithms. In particular, the performance of the proposed scheme
and SG and RLS algorithms is compared with existing techniques,
namely, the full-rank LCMV-SG \cite{frost72} and LCMV-RLS
\cite{romano96}, and the reduced-rank algorithms with
${\boldsymbol S}_D(i)$ designed according to the MSWF
\cite{goldstein}, the AVF \cite{qian} and the optimal linear
beamformer that assumes the knowledge of the covariance matrix
\cite{li}. In particular, the algorithms are compared in terms of
the mean-squared error (MSE) and the
signal-to-interference-plus-noise ratio (SINR), which is defined
for the reduced-rank schemes as
\begin {equation}
\centering {\rm SINR}(i)=\frac{\bar{\boldsymbol
w}^{H}(i){\boldsymbol S}_D^H(i){\boldsymbol R}_{s}{\boldsymbol
S}_D(i) \bar{\boldsymbol w}(i)}{\bar{\boldsymbol
w}^{H}(i){\boldsymbol S}_D^H(i){\boldsymbol R}_{I}{\boldsymbol
S}_D(i)\bar{\boldsymbol w}(i)},
\end{equation}
where ${\boldsymbol R}_{s}$ is the autocorrelation matrix of the
desired signal and ${\boldsymbol R}_{I}$ is the cross-correlation
matrix of the interference and noise in the environment. Note that
for the full-rank schemes the ${\rm SINR}(i)$ assumes
${\boldsymbol S}_D^H(i) = {\boldsymbol I}_M$, where ${\boldsymbol
I}_M$ is an identity matrix with dimensionality $M$. For each
scenario, $200$ runs are used to obtain the curves. In all
simulations, the desired signal power is $\sigma_{d}^{2}=1$, and
the signal-to-noise ratio (SNR) is defined as ${\rm SNR}
=\frac{\sigma_d^2}{\sigma^2}$. The filters are initialized as
$\bar{\boldsymbol w}(0) = [1 ~ 0~ \ldots ~ 0 ]$ and ${\boldsymbol
S}_D(0) = [ {\boldsymbol I}_D^T ~{\boldsymbol 0}_{D \times
(M-D)}^T ]$, where ${\boldsymbol 0}_{D \times M-D}$ is a $D \times
(M-D)$ matrix with zeros in all experiments.

\subsection{MSE Analytical Performance}

In this part of the section, we verify that the results in
(\ref{mre}) and (\ref{fmse}) of the section on MSE convergence
analysis of the proposed reduced-rank SG algorithms can provide a
means of estimating the MSE upon convergence. The steady state MSE
between the desired and the estimated symbol obtained through
simulation is compared with the steady state MSE computed via the
expressions derived in Section VI. In order to illustrate the
usefulness of our analysis we have carried out some experiments.
To semi-analytically compute the MSE for the SG recursion, we have
used (\ref{optf}) and assumed the knowledge of the data covariance
matrix ${\bf R}$. We consider $5$ interferers ($K=6$ users in
total - the SoI and the interferers) at $-60^{o}$, $-30^{o}$,
$0^{o}$, $45^{o}$, $60^{o}$ with powers following a log-normal
distribution with associated standard deviation $3$ dB around the
SoI's power level, which impinges on the array at $15^o$.

We compare the results obtained via simulations with those
obtained by the semi-analytical approach presented in Section VI.
In particular, we consider two sets of parameters in order to
check the validity of our approach. One of the sets has larger
step sizes ($\mu_s=0.0025$ and $\mu_w=0.01$), whereas the other
set employs smaller step sizes ( $\mu_s=0.001$ and $\mu_w=0.001$)
for the recursions. The results shown in Fig. 4 indicate that the
curves obtained with the semi-analytical approach agrees with
those obtained via simulations for both sets of parameters,
verifying the validity of our analysis. Note that the algorithms
with smaller step sizes converge slower than the algorithms
equipped with larger step sizes. However, the proposed algorithms
with smaller step sizes converge to the same level of MSE as the
optimal LCMV, whereas the proposed algorithms with larger step
sizes exhibit a higher level of misadjustment. In what follows, we
will consider the convergence rate of the proposed reduced-rank
algorithms in comparison with existing algorithms.

\subsection{SINR Performance}

In the first two experiments, we consider $7$ interferers at
$-60^{o}$, $-45^{o}$, $- 30^{o}$, $-15^0$, $0^{o}$, $45^{o}$,
$60^{o}$ with powers following a log-normal distribution with
associated standard deviation $3$ dB around the SoI's power level.
The SoI impinges on the array at $30^o$. The parameters of the
algorithms are optimized.

We first evaluate the SINR performance of the analyzed algorithms
against the rank $D$ using optimized parameters ($\mu_s$, $\mu_w$
and forgetting factors $\lambda$) for all schemes and $N=250$
snapshots. The results in Fig. 5 indicate that the best rank for
the proposed scheme is $D=4$ (which will be used in the second
scenario) and it is very close to the optimal full-rank LCMV
filter. Our studies with systems with different sizes show that
$D$ is relatively invariant to the system size, which brings
considerable computational savings. In practice, the rank $D$ can
be adapted in order to obtain fast convergence and ensure good
steady-state performance and tracking after convergence.

We show another scenario in Fig. 6 where the adaptive LCMV filters
are set to converge to the same level of SINR. The parameters used
to obtain these curves are also shown. The SG version of the MSWF
is known to have problems in these situations since it does not
tridiagonalize its covariance matrix \cite{goldstein}, being
unable to approach the optimal LCMV. The curves show an excellent
performance for the proposed scheme which converges much faster
than the full-rank-SG algorithm, and is also better than the more
complex MSWF-RLS and AVF schemes.

In the next experiment, we consider the design of the proposed
adaptive reduced-rank LCMV algorithms equipped with the automatic
rank selection method described in Section V.D. We consider $5$
interferers at $-60^{o}$, $- 30^{o}$, $0^{o}$, $45^{o}$, $60^{o}$
with equal powers to the SoI, which impinges on the array at
$15^o$. Specifically, we evaluate the proposed rank selection
algorithms against the use of fixed ranks, namely, $D=3$ and $D=8$
for both SG and RLS algorithms. The results show that the proposed
automatic rank selection method is capable of ensuring an
excellent trade-off between convergence speed and steady-state
performance, as illustrated in Fig \ref{fig:auto}. In particular,
the proposed algorithm can achieve a significantly faster
convergence performance than the scheme with fixed rank $D=8$,
whereas it attains the same steady state performance.

In the last experiment, we consider a non-stationary scenario
where the system has $6$ users with equal power and the
environment experiences a sudden change at time $i=800$. The $5$
interferers impinge on the ULA at $-60^{o}$, $- 30^{o}$, $0^{o}$,
$45^{o}$, $60^{o}$ with equal powers to the SoI, which impinges on
the array at $15^o$. At time instant $i=800$ we have $3$
interferers with $5$ dB above the SoI's power level entering the
system with DoAs $-45^o$, $-15^o$ and $30^o$, whereas one
interferer with DoA $45^{o}$ and a power level equal to the SoI
exits the system. The proposed and analyzed adaptive beamforming
algorithms are equipped with automatic rank adaptation techniques
and have to adjust their parameters in order to suppress the
interferers. We optimize the step sizes and the forgetting factors
of all the algorithms in order to ensure that they converge as
fast as they can to the same value of SINR. The results of this
experiment are depicted in Fig. \ref{fig:ns}. The curves show that
the proposed reduced-rank algorithms have a superior performance
to the existing algorithms.

\section{Conclusions}

We proposed reduced-rank LCMV beamforming algorithms based on
joint iterative optimization of filters. The proposed reduced-rank
scheme is based on a constrained joint iterative optimization of
filters according to the minimum variance criterion. We derived
LCMV expressions for the design of the projection matrix and the
reduced-rank filter and developed SG and RLS adaptive algorithms
for their efficient implementation along with an automatic rank
selection technique. An analysis of the stability and the
convergence properties of the proposed algorithms was presented
and semi-analytical expressions were derived for predicting the
MSE performance. The numerical results for a digital beamforming
application with a ULA showed that the proposed scheme and
algorithms outperform in convergence and tracking the existing
full-rank and reduced-rank algorithms at comparable complexity.
The proposed algorithms can be extended to other array geometries
and applications .

\begin{appendix}

\section{Preservation of MV and Existence of Multiple Solutions}

In this Appendix we discuss the conditions for which the MV
obtained for the full-rank filter is preserved and the existence
of multiple solutions in the proposed optimization method. Given
an $M \times D$ projection matrix ${\boldsymbol S}_D(i)$, where $D
\leq M$, the ${\rm MV}$ is achieved if and only if ${\boldsymbol
w}$ which minimizes (\ref{flcmv}) belongs to the ${\rm Range}\{
{\boldsymbol S}_D(i) \}$, i.e. ${\boldsymbol w}(i)$ lies in the
subspace generated by ${\boldsymbol S}_D(i)$. In this case, we
have
\begin{equation}
{\rm MV}(\bar{\boldsymbol w}(i)) = ({\boldsymbol
a}^H(\theta_k){\boldsymbol R}^{-1}{\boldsymbol
a}(\theta_k))^{-1}.\end{equation} For a general ${\boldsymbol
S}_D(i)$, we have
\begin{equation}{\rm
MV}(\bar{\boldsymbol w}(i)) \geq ({\boldsymbol
a}^H(\theta_k){\boldsymbol R}^{-1}{\boldsymbol
a}(\theta_k))^{-1}.\end{equation}
From the above relations, we can
conclude that there exists multiple solutions to the proposed
optimization problem. 

\section{Analysis of The Optimization of the Proposed Scheme}

In this appendix, we carry out an analysis of the proposed
reduced-rank method and its optimization. Our approach is based on
expressing the output of the proposed scheme and the proposed
constraint in a convenient form that renders itself to analysis.
Let us rewrite the proposed constrained optimization method in
(\ref{propt}) using the method of Lagrange multipliers and express
it by the Lagrangian
\begin{equation}
\begin{split}
{\mathcal L} & = E\big[ |\bar{\boldsymbol w}^{H}(i){\boldsymbol
S}_D^H(i){\boldsymbol r}(i)|^2 \big]  + 2\Re [\lambda
(\bar{\boldsymbol w}^H(i){\boldsymbol S}_D^H(i){\boldsymbol
a}(\theta_k)-1) ] 
, \label{uopt2}
\end{split}
\end{equation}
In order to proceed, let us express $\bar{x}(i)$ in an alternative
and more convenient form as
\begin{equation}
\begin{split}
\bar{x}(i) &  = \bar{\boldsymbol w}^H(i) {\boldsymbol S}_D^H(i)
{\boldsymbol r}(i) =  \bar{\boldsymbol w}^H(i) \sum_{d=1}^D
{\boldsymbol s}_d^H(i) {\boldsymbol r}(i) {\boldsymbol q}_d \\ & =
\bar{\boldsymbol w}^H(i) \left[\begin{array}{ccccc} {\boldsymbol
r}(i) & 0 & 0 & \ldots & 0 \\ 0 & {\boldsymbol r}(i) & 0 & \ldots
& 0 \\ \vdots & \vdots & \vdots & \ddots & \vdots \\ 0 & \ldots &
0 & 0 & {\boldsymbol r}(i) \end{array} \right]^T
\left[\begin{array}{c}
{\boldsymbol s}_1^*(i) \\ {\boldsymbol s}_2^*(i) \\ \vdots \\
{\boldsymbol s}_D^*(i)
\end{array} \right] \\ & = \bar{\boldsymbol w}^H(i) {\boldsymbol
\Re}^T(i) {\boldsymbol s}_v^*(i)
\end{split}
\end{equation}
where ${\boldsymbol \Re}(i)$ is a $DM \times D$ block diagonal
matrix with the input data vector ${\boldsymbol r}(i)$,
${\boldsymbol q}_d$ is a $D \times 1$ vector with a $1$ in the
$d$-th position and ${\boldsymbol s}_v^*(i)$ is a $DM \times 1$
vector with the columns of ${\boldsymbol S}_D(i)$ stacked on top
of each other.

In order to analyze the proposed joint optimization procedure, we
can rearrange the terms in $\bar{x}(i)$ and define a single
$D(M+1) \times 1$ parameter vector ${\boldsymbol f}(i) =
[\bar{\boldsymbol w}^T(i) ~ {\boldsymbol s}_v^T(i)]^T$. We can
therefore further express $\bar{x}(i)$ as
\begin{equation}
\begin{split}
\bar{x}(i) &  = {\boldsymbol f}^H(i) \left[\begin{array}{cc}
{\boldsymbol 0}_{D \times D} & {\boldsymbol 0}_{D \times DM} \\
{\boldsymbol \Re}(i) & {\boldsymbol 0}_{DM \times DM}
\end{array} \right] {\boldsymbol f}(i) \\ & = {\boldsymbol f}^H(i)
{\boldsymbol G}(i) {\boldsymbol f}(i)
\end{split}
\end{equation}
where ${\boldsymbol G}(i)$ is a $D(M+1) \times D(M+1)$ matrix
which contains ${\boldsymbol \Re}(i)$. Now let us perform a
similar linear algebra transformation with the proposed constraint
$\bar{\boldsymbol w}^H(i) {\boldsymbol S}_D^H(i) {\boldsymbol
a}(\theta_k) = 1$ and express it as
\begin{equation}
\bar{\boldsymbol w}^H(i) {\boldsymbol S}_D^H(i) {\boldsymbol
a}(\theta_k) = {\boldsymbol f}^H(i) {\boldsymbol A}(\theta_k)
{\boldsymbol f}(i)
\end{equation}
where the $D(M+1) \times D(M+1)$ matrix ${\boldsymbol
A}(\theta_k)$ is structured as
\begin{equation}
{\boldsymbol A}(\theta_k) = \left[\begin{array}{cc}
{\boldsymbol 0}_{D \times D} & {\boldsymbol 0}_{D \times DM} \\
{\boldsymbol \Re}_{\boldsymbol a(\theta_k)} & {\boldsymbol 0}_{DM
\times DM}
\end{array} \right]
\nonumber
\end{equation}
and the $DM \times D$ block diagonal matrix ${\boldsymbol
\Re}_{\boldsymbol a(\theta_k)}(i)$ with the steering vector
${\boldsymbol a}(\theta_k)$ constructed as
\begin{equation}
{\boldsymbol \Re}_{\boldsymbol a(\theta_k)} =
\left[\begin{array}{ccccc} {\boldsymbol a}(\theta_k) & 0 & 0 &
\ldots & 0
\\ 0 & {\boldsymbol a}(\theta_k) & 0 & \ldots & 0 \\ \vdots & \vdots &
\vdots & \ddots & \vdots \\ 0 & \ldots & 0 & 0 & {\boldsymbol
a}(\theta_k) \end{array} \right]
\end{equation}
At this point, we can alternatively express the Lagrangian in
(\ref{uopt2}) as
\begin{equation}
\begin{split}
{\mathcal L} & = E\big[ |{\boldsymbol f}^H(i) {\boldsymbol G}(i)
{\boldsymbol f}(i)|^2 \big]  + 2\Re [\lambda ({\boldsymbol f}^H(i)
{\boldsymbol A}(\theta_k) {\boldsymbol f}(i)-1) ]. \label{uopt3}
\end{split}
\end{equation}
We can examine the convexity of the above Lagrangian by computing
the Hessian (${\boldsymbol H}$)with respect to ${\boldsymbol
f}(i)$ using the expression \cite{bert}
\begin{equation}
{\boldsymbol H} = \frac{\partial}{\partial {\boldsymbol f}^H(i)}
\frac{\partial ({\mathcal L})}{\partial {\boldsymbol f}(i)}
\label{hess}
\end{equation}
and testing if the terms are positive semi-definite. Specifically,
${\boldsymbol H}$ is positive semi-definite if ${\boldsymbol
v}^{H}{\boldsymbol H}{\boldsymbol v} \geq 0$ for all nonzero
${\boldsymbol v} \in \boldsymbol{C}^{D(M+1)\times D(M+1)}$
\cite{golub}. Therefore, the optimization problem is convex if the
Hessian ${\boldsymbol H}$ is positive semi-definite.

Evaluating the partial differentiation in the expression given in
(\ref{hess}) yields
\begin{equation}
\begin{split}
{\boldsymbol H} & = E\big[ {\boldsymbol f}^H(i) {\boldsymbol G}(i)
{\boldsymbol f}(i) {\boldsymbol G}(i)  +  {\boldsymbol
G}(i) {\boldsymbol f}(i) {\boldsymbol f}^H(i){\boldsymbol G}(i) \\
& \quad +  {\boldsymbol G}(i) {\boldsymbol f}^H(i){\boldsymbol
G}(i) {\boldsymbol f}(i)  +  {\boldsymbol f}^H(i){\boldsymbol
G}(i)  {\boldsymbol G}(i){\boldsymbol f}(i)  +  2 \lambda
{\boldsymbol A}(\theta_k) \big]
\end{split}
\end{equation}
By examining ${\boldsymbol H}$, we verify that the second and
fourth terms are positive semi-definite, whereas the first and the
third terms are indefinite. The fifth term depends on the
constraint, which is typically positive in the proposed scheme as
verified in our studies, yielding a positive semi-definite matrix.
Therefore, the optimization problem can not be classified as
convex. It is however important to remark that our studies
indicate that there are no local minima and there exists multiple
solutions (which are possibly identical).

In order to support this claim, we have checked the impact on the
proposed algorithms of different initializations . This study
confirmed that the algorithms are not subject to performance
degradation due to the initialization although we have to bear in
mind that the initialization ${\boldsymbol S}_D(0)= {\boldsymbol
0}_{M \times D}$ annihilates the signal and must be avoided. We
have also studied a particular case of the proposed scheme when
$M=1$ and $D=1$, which yields the Lagrangian $ {\mathcal
L}(\bar{\boldsymbol w}, {\boldsymbol S}_D) = E \big[ | \bar{w} S_D
r |^2 \big] + 2\Re \big[ \lambda (\bar{w} S_D a(\theta_k) -1)
\big]$. Choosing $S_D$ (the "scalar" projection) fixed with $D$
equal to $1$, it is evident that the resulting function ${\mathcal
L}(\bar{w},S_D=1,r) = |w^*~r|^{2}+ 2\Re \big[ \lambda (\bar{w}
a(\theta_k) -1) \big]$ is a convex one. In contrast to that, for a
time-varying projection $S_D$ the plots of the function indicate
that the function is no longer convex but it also does not exhibit
local minima. This problem can be generalized to the vector case,
however, we can no longer verify the existence of local minima due
to the multi-dimensional surface. This remains as an interesting
open problem.

\section{Derivation and Simplification of ${\boldsymbol S}_D(i)$}

In this appendix, we detail the derivation of the filter
${\boldsymbol S}_D(i)$ and the simplification shown in (\ref{18})
for reducing the computational complexity. Let us consider the
derivation of ${\boldsymbol S}_D(i)$ obtained from the
minimization of the Lagrangian
\begin{equation}
\begin{split}
{\mathcal L}({\boldsymbol S}_D(i), \bar{\boldsymbol w}(i)) & =
\sum_{l=1}^{i} \alpha^{i-l} |\bar{\boldsymbol
w}^{H}(i){\boldsymbol S}_D^H(i){\boldsymbol r}(l)|^2   + 2\Re
[\lambda (\bar{\boldsymbol w}^H(i){\bf S}_D^H(i){\boldsymbol
a}(\theta_k)-1) ], \label{uopt2}
\end{split}
\end{equation}
Taking the gradient terms of the above expression with respect to
${\boldsymbol S}_D^*(i)$, we get
\begin{equation}
\begin{split}
\nabla {\mathcal L}({\boldsymbol S}_D(i), \bar{\boldsymbol
w}(i))_{{\boldsymbol S}_D^*(i)} & = \sum_{l=1}^{i} \alpha^{i-l}
{\boldsymbol r}(l) {\boldsymbol r}^H(l) {\boldsymbol S}_D(i)
\bar{\boldsymbol w}(i) \bar{\boldsymbol w}^H(i) + 2 \lambda
{\boldsymbol a}(\theta_k) \bar{\boldsymbol w}^H(i) \\ & =
{\boldsymbol R}(i) {\boldsymbol S}_D(i) \bar{\boldsymbol
R}_{\bar{w}}(i) + 2 \lambda {\boldsymbol a}(\theta_k)
\bar{\boldsymbol w}^H(i).
\end{split}
\end{equation}
Making the above gradient terms equal to zero yields
\begin{equation}
{\boldsymbol S}_D(i) = {\boldsymbol R}^{-1}(i) (-2 \lambda)
{\boldsymbol a}(\theta_k) \bar{\boldsymbol w}^H(i)
\bar{\boldsymbol R}_{\bar{w}}^{-1}. \label{filtsdwl}
\end{equation}
Using the proposed constraint $\bar{\boldsymbol
w}^H(i){\boldsymbol S}_D^H(i) {\boldsymbol a}(\theta_k) =1$ and
substituting the above filter expression, we obtain the Lagrange
multiplier $\lambda = -1/2 (\bar{\boldsymbol w}^H(i)
\bar{\boldsymbol R}_{\bar{w}}^{-1} \bar{\boldsymbol w}(i)
{\boldsymbol a}^H(\theta_k) {\boldsymbol R}^{-1}(i) {\boldsymbol
a}(\theta_k))^{-1}$. Substituting $\lambda$ into (\ref{filtsdwl}),
we get
\begin{equation}\label{filtsd2}
\boldsymbol S_{D}(i)=\frac{\boldsymbol R^{-1}(i)\boldsymbol
a(\theta_k)\bar{\boldsymbol w}^{H}(i)\bar{\boldsymbol
R}_{\bar{w}}^{-1}(i)}{\bar{\boldsymbol w}^{H}(i)\bar{\boldsymbol
R}_{\bar{w}}^{-1}(i)\bar{\boldsymbol w}(i){\boldsymbol
a}^{H}(\theta_k) { \boldsymbol R}^{-1}(i) {\boldsymbol
a}(\theta_k)}
\end{equation}
The above expression for the matrix filter ${\boldsymbol S}_D(i)$
can be simplified by observing the quantities involved and making
use of the proposed constraint $\bar{\boldsymbol w}^{H}(i) {
\boldsymbol S}_D^{H}(i) {\boldsymbol a}(\theta_k)=1$. Let us
consider the term $\bar{\boldsymbol w}^H(i) \bar{\boldsymbol
R}_{\bar{w}}^{-1} \bar{\boldsymbol w}(i)$ in the denominator of
(\ref{filtsd2}) and multiply it by the proposed constraint as
follows:
\begin{equation}
\begin{split}
\bar{\boldsymbol w}^H(i) \bar{\boldsymbol R}_{\bar{w}}^{-1}
\bar{\boldsymbol w}(i) & =  \bar{\boldsymbol w}^H(i)
\bar{\boldsymbol R}_{\bar{w}}^{-1} \bar{\boldsymbol w}(i)
\bar{\boldsymbol w}^{H}(i) {\boldsymbol S}_D^{H}(i) {\boldsymbol
a}(\theta_k) \\ & = \bar{\boldsymbol w}^{H}(i) {\boldsymbol
S}_D^{H}(i) {\boldsymbol a}(\theta_k) =  1. \label{wrw}
\end{split}
\end{equation}
Now let us consider the term ${\boldsymbol a}^{H}(\theta_k)
\bar{\boldsymbol w}^H(i)  \bar{\boldsymbol R}^{-1}_{\bar{w}}(i)$
and rewrite it as follows:
\begin{equation}
\begin{split}
{\boldsymbol a}(\theta_k) \bar{\boldsymbol w}^H(i)
\bar{\boldsymbol R}^{-1}_{\bar{w}}(i) & = {\boldsymbol
a}(\theta_k) \bar{\boldsymbol w}^H(i) \bar{\boldsymbol
R}^{-1}_{\bar{w}}(i) \bar{\boldsymbol w}^{H}(i) {\boldsymbol
S}_D^{H}(i) {\boldsymbol a}(\theta_k) \\ & = {\boldsymbol
a}(\theta_k){\boldsymbol a}^{H}(\theta_k)   {\boldsymbol
S}_D(i)\bar{\boldsymbol w}(i) \bar{\boldsymbol w}^H(i)
\bar{\boldsymbol R}^{-1}_{\bar{w}}(i)
\\ & = {\boldsymbol
a}(\theta_k) {\boldsymbol a}^{H}(\theta_k)   {\boldsymbol S}_D(i)
 = {\boldsymbol a}(\theta_k) \bar{\boldsymbol a}^{H}(\theta_k)
 .\label{rw}
\end{split}
\end{equation}
Using the relations obtained in (\ref{wrw}) and (\ref{rw}) into
the expression in (\ref{filtsd2}), we can get a simpler expression
for the projection matrix as given by
\begin{equation}
\begin{split}
\label{filtsd3} \boldsymbol S_{D}(i) & =\frac{{\boldsymbol
R}^{-1}(i) {\boldsymbol a}(\theta_k)\bar{\boldsymbol
w}^{H}(i)\bar{\boldsymbol R}_{\bar{w}}^{-1}(i)}{\bar{\boldsymbol
w}^{H}(i)\bar{\boldsymbol R}_{\bar{w}}^{-1}(i)\bar{\boldsymbol
w}(i){\boldsymbol a}^{H}(\theta_k) {\boldsymbol R}^{-1}(i)
{\boldsymbol a}(\theta_k)} = \frac{ {\boldsymbol R}^{-1}(i)
\overbrace{ {\boldsymbol a}(\theta_k) \bar{\boldsymbol
w}^{H}(i)\bar{\boldsymbol R}_{\bar{w}}^{-1}(i)}^{{\boldsymbol
a}(\theta_k) \bar{\boldsymbol a}^{H}(\theta_k)}}{ \underbrace{
\bar{\boldsymbol w}^{H}(i)\bar{\boldsymbol
R}_{\bar{w}}^{-1}(i)\bar{\boldsymbol w}(i)}_{1}{\boldsymbol
a}^{H}(\theta_k) { \boldsymbol R}^{-1}(i) {\boldsymbol
a}(\theta_k)} \\ & = \frac{{\boldsymbol R}^{-1}(i){\boldsymbol
a}(\theta_k)  \bar{\boldsymbol a}^H(\theta_k)}{ {\boldsymbol
a}^{H}(\theta_k) { \boldsymbol R}^{-1}(i) {\boldsymbol
a}(\theta_k)}
\end{split}
\end{equation}
This completes the derivation and the simplification.

\end{appendix}

\newpage

\begin{figure}[!htb]
\begin{center}
\def\epsfsize#1#2{0.5\columnwidth}
\epsfbox{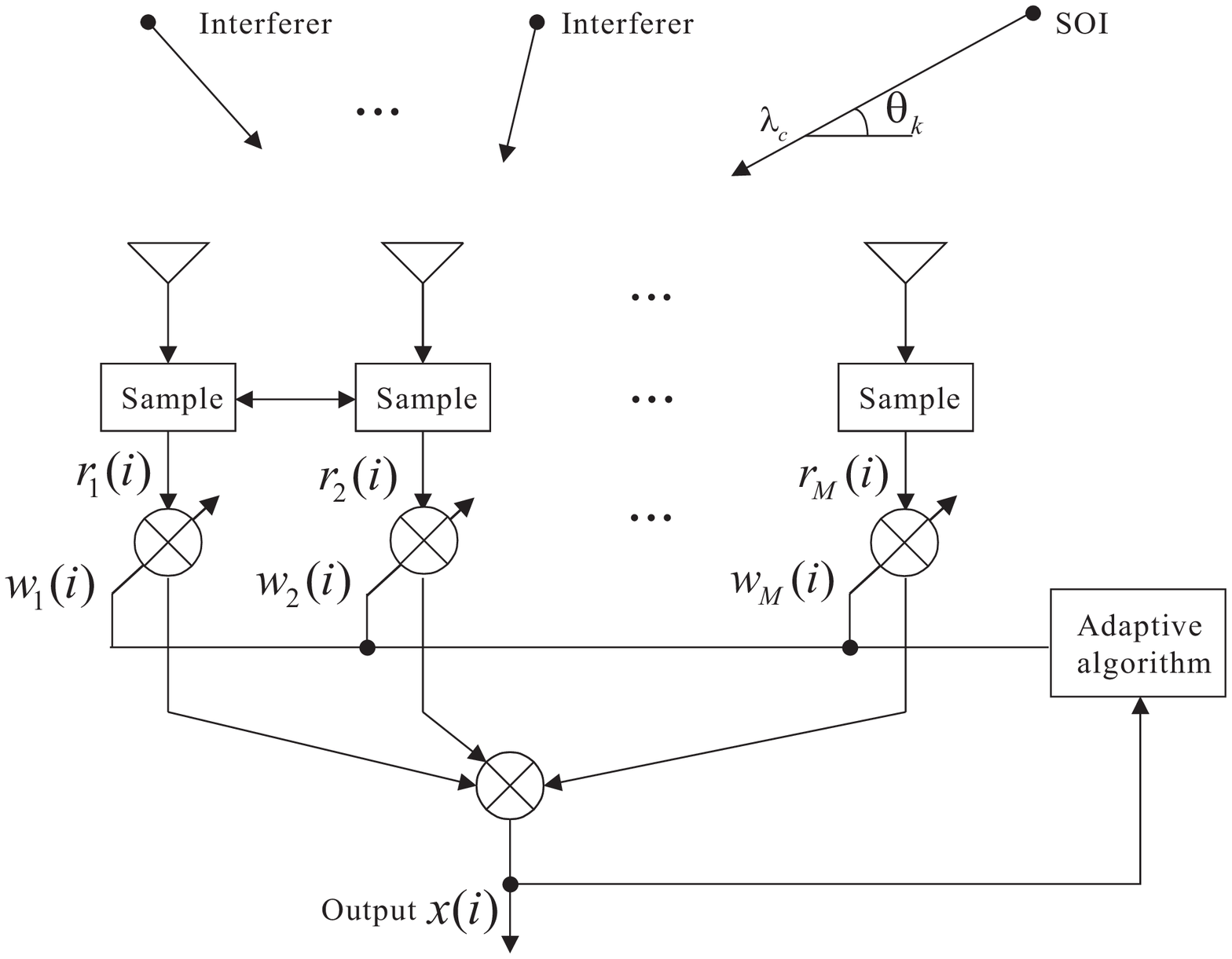} \vspace*{-1em}\caption{Schematic of a linear
antenna array system with interferers.}
\end{center}
\end{figure}

\begin{figure}[h]
       \vspace*{-1.0em}\centering  
       \hspace*{-3.85em}{\includegraphics[width=11.0cm, height=4.5cm]{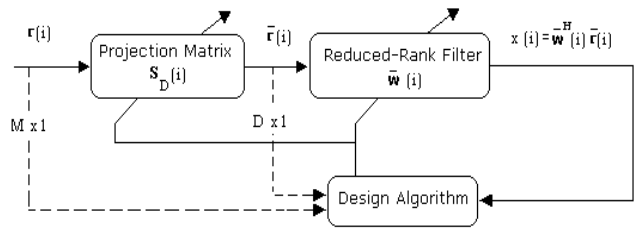}}
        \vspace*{-3em}
       \caption{Schematic of the proposed reduced-rank scheme.}
 \label{SchemeSetup}
\end{figure}

\begin{table}[h]
\centering%
\caption{\small Computational complexity of LCMV algorithms.} {
\begin{tabular}{lcc}
\hline
{\small Algorithm} & {\small Additions} & {\small Multiplications} \\
\hline \emph{\small \bf Full-rank-SG }\cite{frost72}  & {\small
$3M+1$} & {\small $3M+2$} \\ \\
\emph{\small \bf Full-rank-RLS }\cite{romano96} & {\small $3M^2 -
2M + 3$} & {\small $6M^{2}+2M + 2$}
\\ \\
\emph{\small \bf Proposed-SG} \cite{delamareelb}  & {\small $3DM + 2M $}  & {\small $ 3DM + M $}  \\
\emph{\small \bf }  & {\small $ +2D-2$}  & {\small $  +5D + 2$}  \\
\emph{\small \bf Proposed-RLS}  & {\small $ 3M^2 - 2M + 3 $}  & {\small $ 7M^2 + 2M $}  \\
\emph{\small \bf }  & {\small $+3D^2 - 8D + 3$}  & {\small $ +7D^2+9D$}  \\
\emph{\small \bf MSWF-SG }\cite{goldstein} & {\small $DM^2 - M^2
$} & {\small $DM^2 - M^2 $}  \\ \emph{\small \bf
} & {\small $+3D -2$} & {\small $ +2DM +4D + 1$}  \\
\emph{\small \bf MSWF-RLS }\cite{goldstein} & {\small $DM^2 + M^2
+ 6D^2 $} & {\small $DM^2+M^2$} \\
\emph{\small } & {\small $-8D+2$} & {\small $+2DM+3D+2$} \\
\emph{\small \bf AVF }\cite{qian} & {\small $D((M)^2+3(M-1)^2)-1$} & {\small $D(4M^2+4M + 1)$} \\
\emph{\small \bf  } & {\small $+D(5(M-1)+1)+2M$} & {\small $+4M +
2$}
\\\hline
\end{tabular}
}
\end{table}

\begin{figure}[!htb]
\begin{center}
\def\epsfsize#1#2{0.6\columnwidth}
\epsfbox{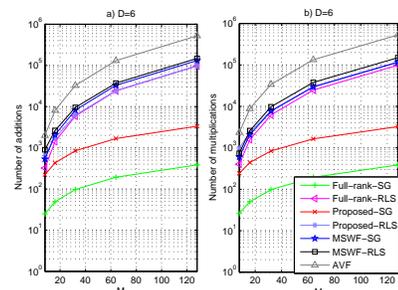} \vspace*{-1.5em} \caption{Complexity in
terms of arithmetic operations against $M$. }
\end{center}
\end{figure}

\begin{figure}[!htb]
\begin{center}
\def\epsfsize#1#2{0.625\columnwidth}
\epsfbox{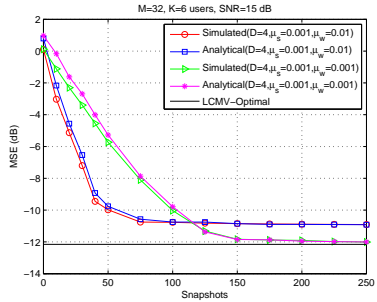} \vspace*{-1em}\caption{\small MSE
analytical versus simulated performance for the proposed
reduced-rank SG algorithm.}
\end{center}
\end{figure}

\begin{figure}[!htb]
\begin{center}
\def\epsfsize#1#2{0.625\columnwidth}
\epsfbox{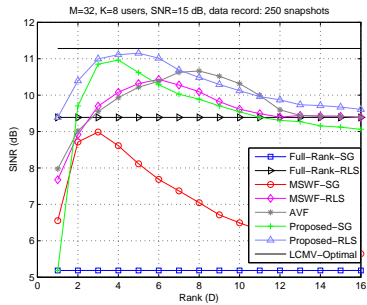} \vspace*{-1em}\caption{\small SINR
performance of LCMV algorithms against rank ($D$) with $M=32$,
$SNR=15$ dB, $N= 250$ snapshots.}
\end{center}
\end{figure}

\begin{figure}[!htb]
\begin{center}
\def\epsfsize#1#2{0.625\columnwidth}
\epsfbox{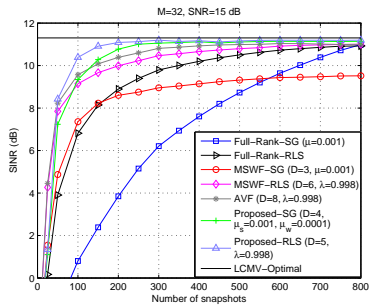} \vspace*{-1em}\caption{\small
SINR performance of LCMV algorithms against snapshots with $M=32$,
$SNR=15$ dB.}
\end{center}
\end{figure}

\begin{figure}[!htb]
\begin{center}
\def\epsfsize#1#2{0.625\columnwidth}
\epsfbox{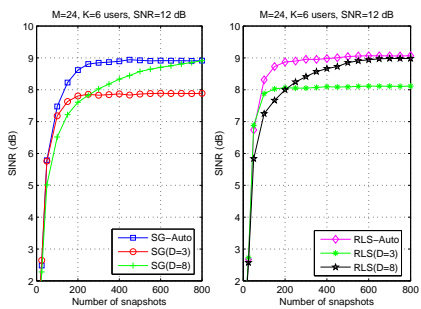} \vspace*{-1em} \caption{SINR
performance of LCMV (a) SG and (b) RLS algorithms against
snapshots with $M=24$, $SNR=12$ dB with automatic rank selection.}
\label{fig:auto}
\end{center}
\end{figure}

\begin{figure}[!htb]
\begin{center}
\def\epsfsize#1#2{0.625\columnwidth}
\epsfbox{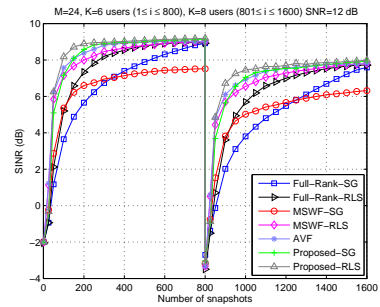} \vspace*{-1em} \caption{SINR
performance of LCMV algorithms against snapshots with $M=24$,
$SNR=12$ dB in a non-stationary scenario.} \label{fig:ns}
\end{center}
\end{figure}

\end{document}